\documentclass[aps, superscriptaddress, longbibliography, notitlepage, twocolumn, floatfix, nofootinbib]{revtex4-2}     

\usepackage{amsmath,amssymb}
\usepackage{graphicx}
\usepackage{enumitem}
\usepackage{mathtools}
\usepackage{color}
\usepackage{array} 
\usepackage{tabularx} 

\usepackage[colorlinks=true, urlcolor=blue, anchorcolor=blue, citecolor=blue, linkcolor=blue]{hyperref}

\begin{document}

\title{$(\omega_1, \omega_2)$-temporal random hyperbolic graphs}

\author{Sofoclis Zambirinis}
\affiliation{Department of Electrical Engineering, Computer Engineering and Informatics, Cyprus University of Technology, 3036 Limassol, Cyprus}
\author{Fragkiskos Papadopoulos}
\email{f.papadopoulos@cut.ac.cy}
\affiliation{Department of Electrical Engineering, Computer Engineering and Informatics, Cyprus University of Technology, 3036 Limassol, Cyprus}

\date{\today}

\begin{abstract}
We extend a recent model of temporal random hyperbolic graphs by allowing connections and disconnections to persist across network snapshots with different probabilities, $\omega_1$ and $\omega_2$. This extension, while conceptually simple, poses analytical challenges involving the Appell $F_1$ series. Despite these challenges, we are able to analyze key properties of the model, which include the distributions of contact and intercontact durations, as well as the expected time-aggregated degree. The incorporation of $\omega_1$ and $\omega_2$ enables more flexible tuning of the average contact and intercontact durations, and of the average time-aggregated degree, providing a finer control for exploring the effect of temporal network dynamics on dynamical processes. Overall, our results provide new insights into the analysis of temporal networks and contribute to a more general representation of real-world scenarios.
\end{abstract}

\maketitle

%%%%%%%%%%%%%%%%%%%%%%%%%%%%%%%%%%%%%%%%%%%%%%%%%%%%%%%%%%%%%%%%%%%%%%%%%%%%%%%%%%%%%%%%%%%%%%%%%%%%%%%%%%%%%%%%%%%%%%
\section{Introduction}

Originally motivated by the parsimonious modeling of human contact networks~\cite{Starnini2013, StarniniDevices2017, flores2018}, a simple model of temporal random hyperbolic graphs has been recently introduced and analyzed, called dynamic-$\mathbb{S}^1$~\cite{Papadopoulos2019}. The model has demonstrated the ability to qualitatively and sometimes quantitatively reproduce various dynamical properties observed in real temporal networks. These properties include broad distributions of contact and intercontact durations, broad weight and strength distributions, narrow distributions of shortest time-respecting paths, and formation of recurrent components~\cite{Papadopoulos2019}. In the model, each node is endowed with an expected degree or popularity variable $\kappa$ and a similarity coordinate $\theta$. Each network snapshot is then independently generated according to the $\mathbb{S}^1$ model, or equivalently, the hyperbolic $\mathbb{H}^2$ model~\cite{Krioukov2010}, where nodes connect with probability $p(\chi)=1/(1+\chi^{1/T})$. Here, $\chi\propto \Delta\theta/(\kappa \kappa')$ represents the effective distance between the nodes, $\Delta\theta$ is the nodes' angular similarity distance, $\kappa$ and $\kappa'$ are the nodes' expected degrees, and parameter $T \in (0,1)$ is called network temperature. We note that the dynamic-$\mathbb{S}^1$ yields realistic dynamical properties only for $T \in (0,1)$, but not for $T > 1$~\cite{Papadopoulos2022}.

While the snapshots are independently generated in the dynamic-$\mathbb{S}^1$, they are not independent as there are correlations among them induced by the nodes' effective distances. For instance, nodes at smaller effective distances have higher chances of being connected in consecutive snapshots. Given the ability of the model to adequately reproduce various dynamical properties of real systems, it has been demonstrated that spreading processes perform remarkably similar in some real networks and their modeled counterparts~\cite{Papadopoulos2019}. Furthermore, the model has already demonstrated its utility in real-world epidemiological studies~\cite{Andrianou2022}, and has been employed to justify the meaningful mapping of human proximity networks into hyperbolic spaces~\cite{flores2020map}.

To better capture the average contact and intercontact durations observed in some real systems, the dynamic-$\mathbb{S}^1$ has been recently extended to account for \emph{link persistence}, where connections and disconnections can persist, i.e., propagate, from one snapshot to the next, irrespective of their effective distance~\cite{mazzarisi2020, Papadopoulos2019lp, hartle2021, friel2016}. This extension, called $\omega$-dynamic-$\mathbb{S}^1$~\cite{Zambirinis2022}, introduces the probability parameter $\omega \in [0,1)$, dictating the persistence of both connections and disconnections. 

However, the assumption that links and non-links persist with the same probability may not generally hold in reality. For instance, consider collaboration networks. Here, if two nodes (e.g., authors) collaborate at least once, then a link between them will always exist in the network. However, this does not imply that two existing nodes that have never collaborated will never do so in the future. As another example, consider connected Internet Service Providers (ISPs) separated by large geographic distances. Such connections are expected to persist as they are generally expensive to establish. On the other hand, disconnected ISPs at small geographic distances may not remain disconnected with equally high probability, as the costs and logistical barriers of such connections can be significantly lower. In general, different factors can affect the persistence of connections and disconnections depending on the context. Moreover, by using a common persistence probability for links and non-links, the $\omega$-dynamic-$\mathbb{S}^1$ does not allow individual tuning of the average contact and intercontact durations, as both are dictated by the same parameter $\omega$.

To address these limitations, here we generalize the model by allowing connections and disconnections to persist with different probabilities, denoted as $\omega_1$ and $\omega_2$. We refer to the generalized model as ($\omega_1, \omega_2$)-dynamic-$\mathbb{S}^1$. Even though this generalization is conceptually simple, it poses significant analytical challenges involving the Appell $F_1$ series---a two-variable generalization of the Gauss hypergeometric function~\cite{special_functions_book}. In our case, these variables involve the persistence probabilities $\omega_1$ and $\omega_2$. In contrast, the analysis simplifies if $\omega_1=\omega_2$, requiring only manipulations with the Gauss hypergeometric function~\cite{Zambirinis2022}. 

\begin{figure*}[t]
\includegraphics[width=2.3in]{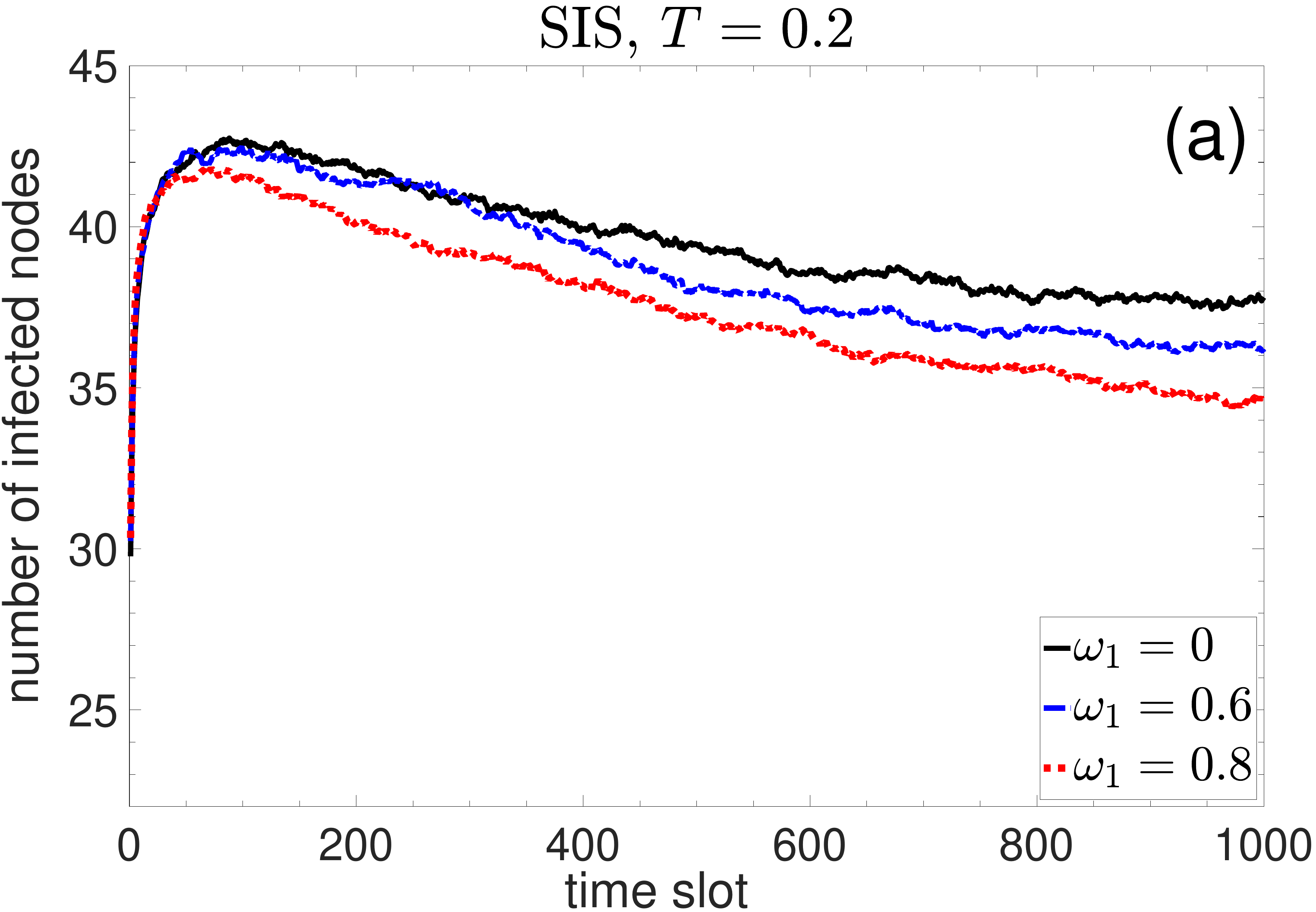}
\includegraphics[width=2.3in]{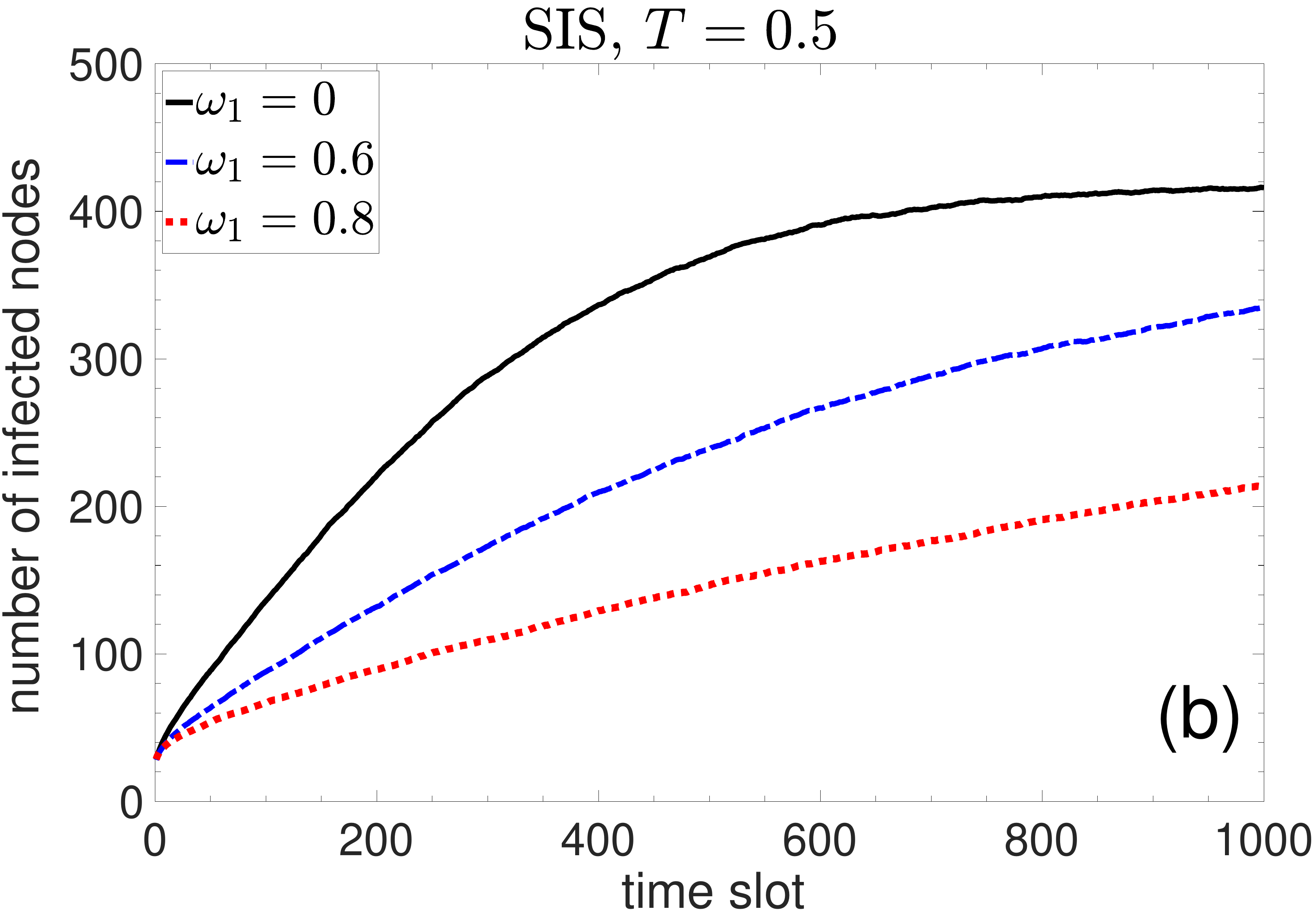}
\includegraphics[width=2.3in]{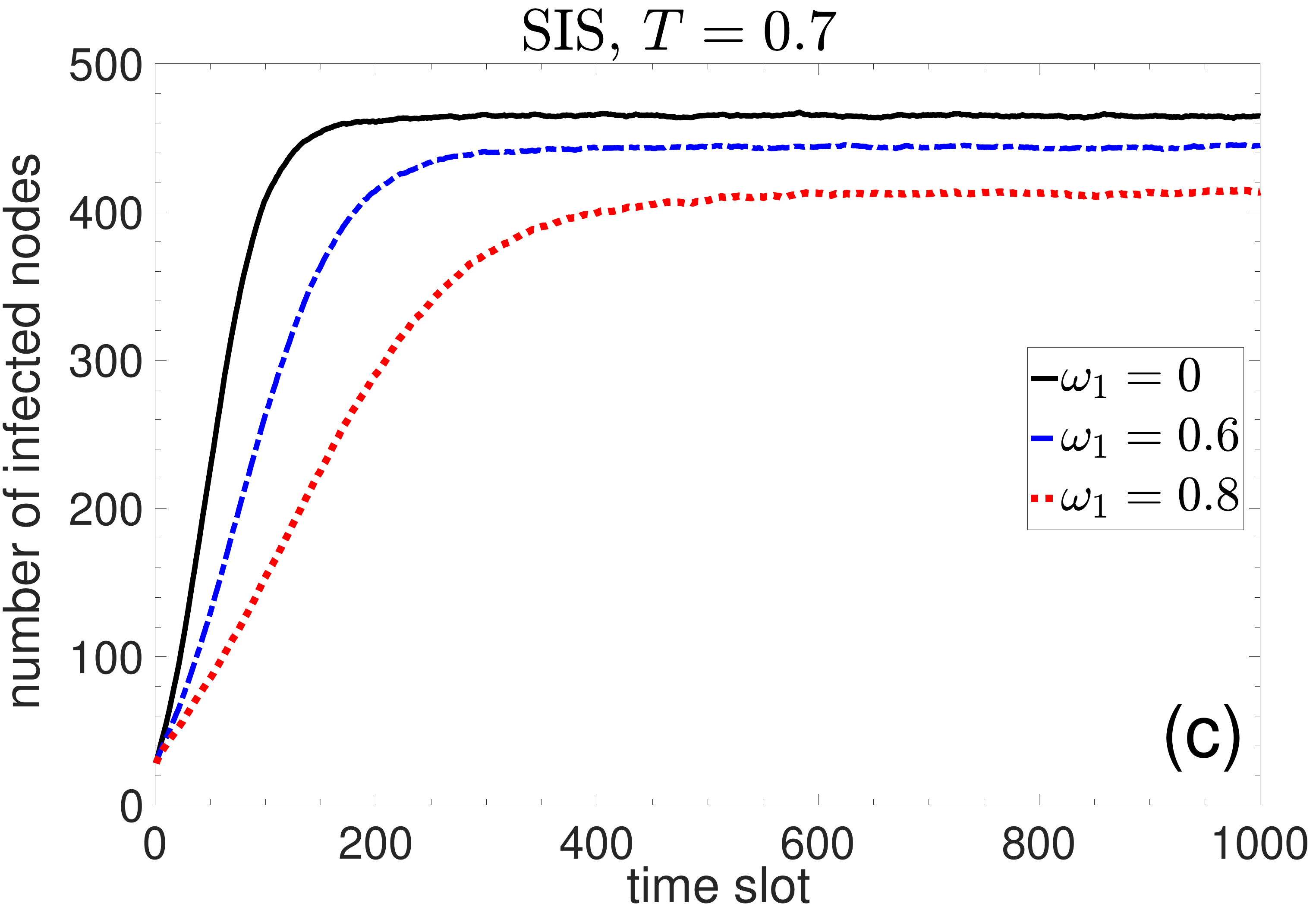}
\newline
\newline
\includegraphics[width=2.3in]{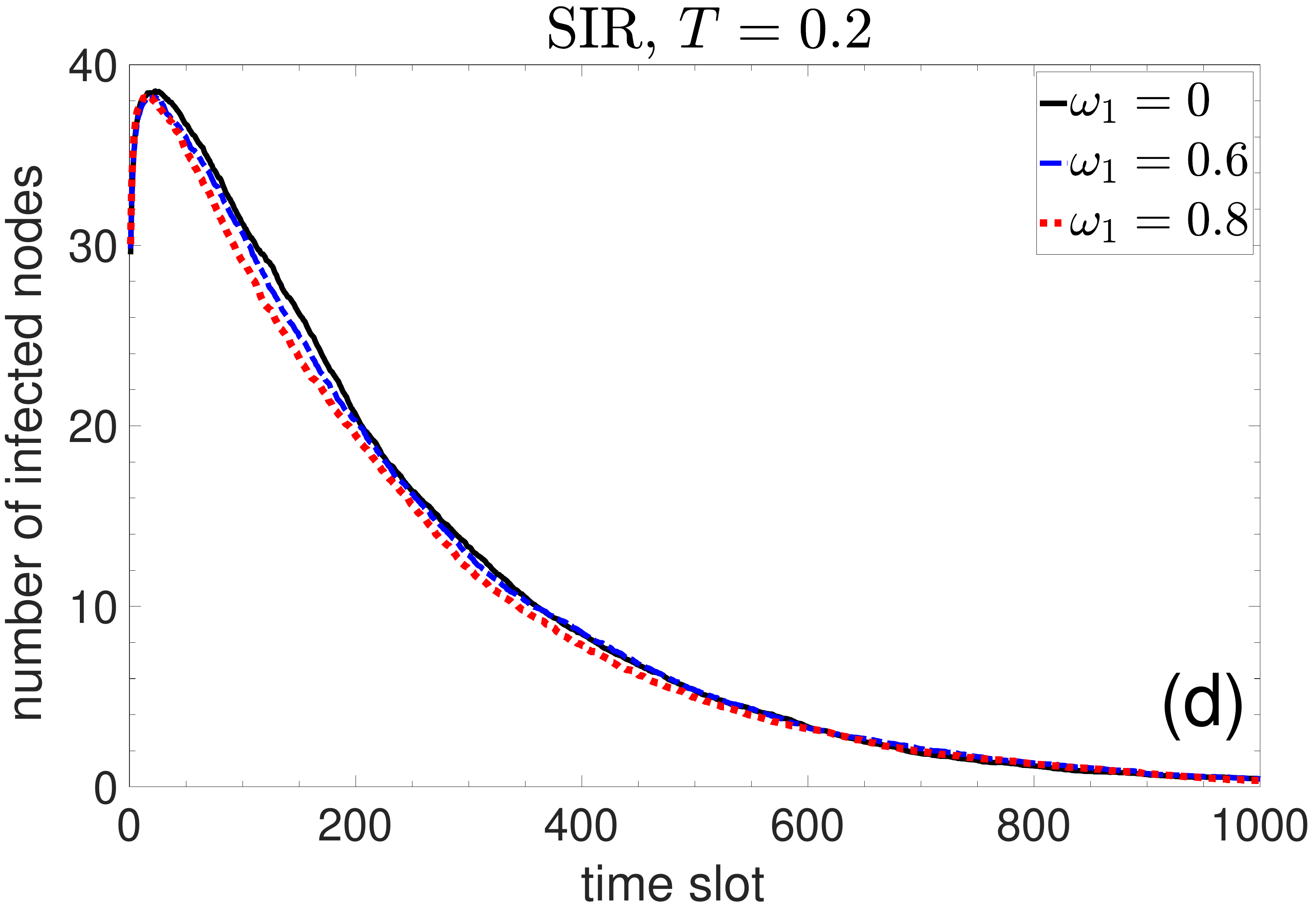}
\includegraphics[width=2.3in]{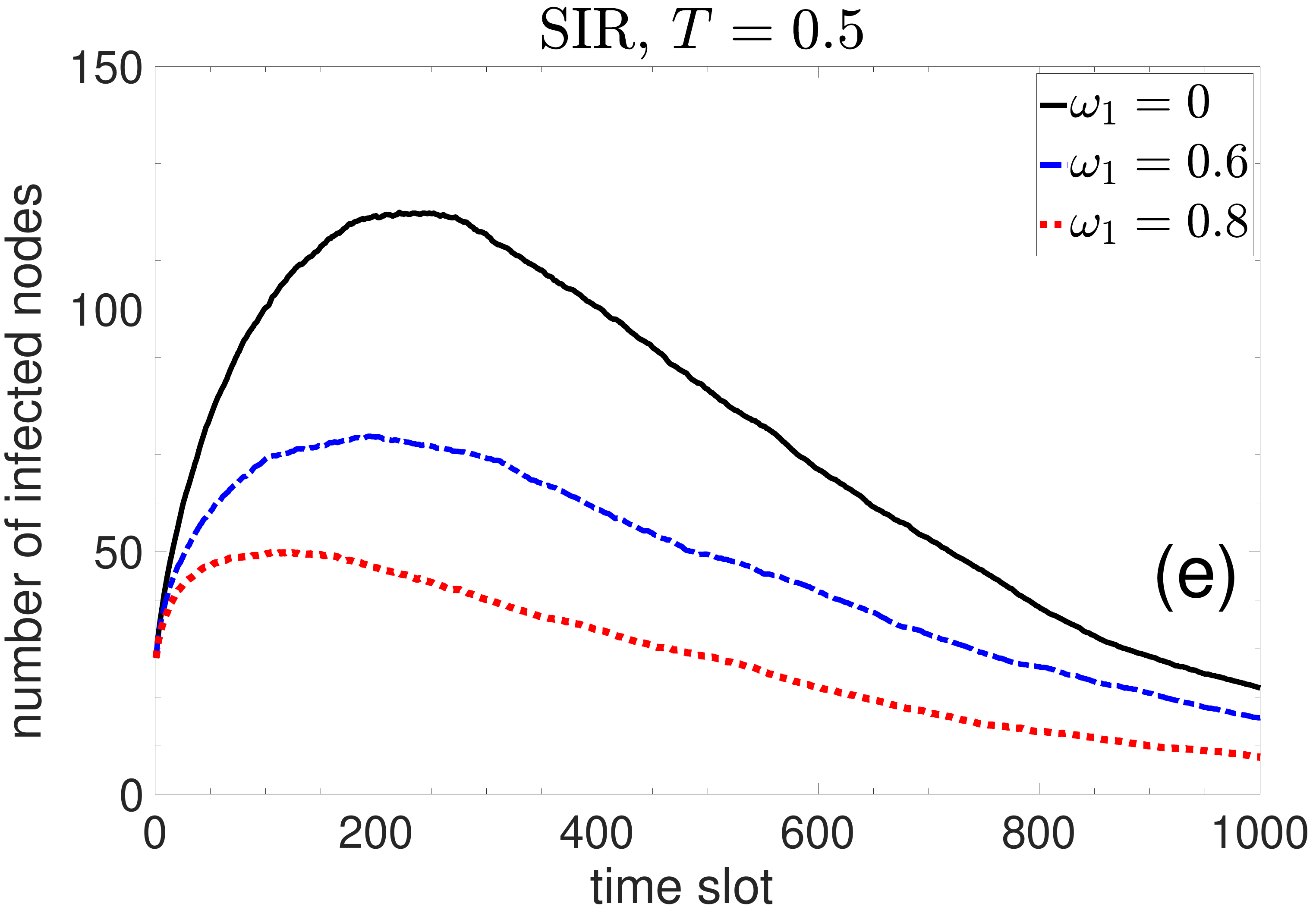}
\includegraphics[width=2.3in]{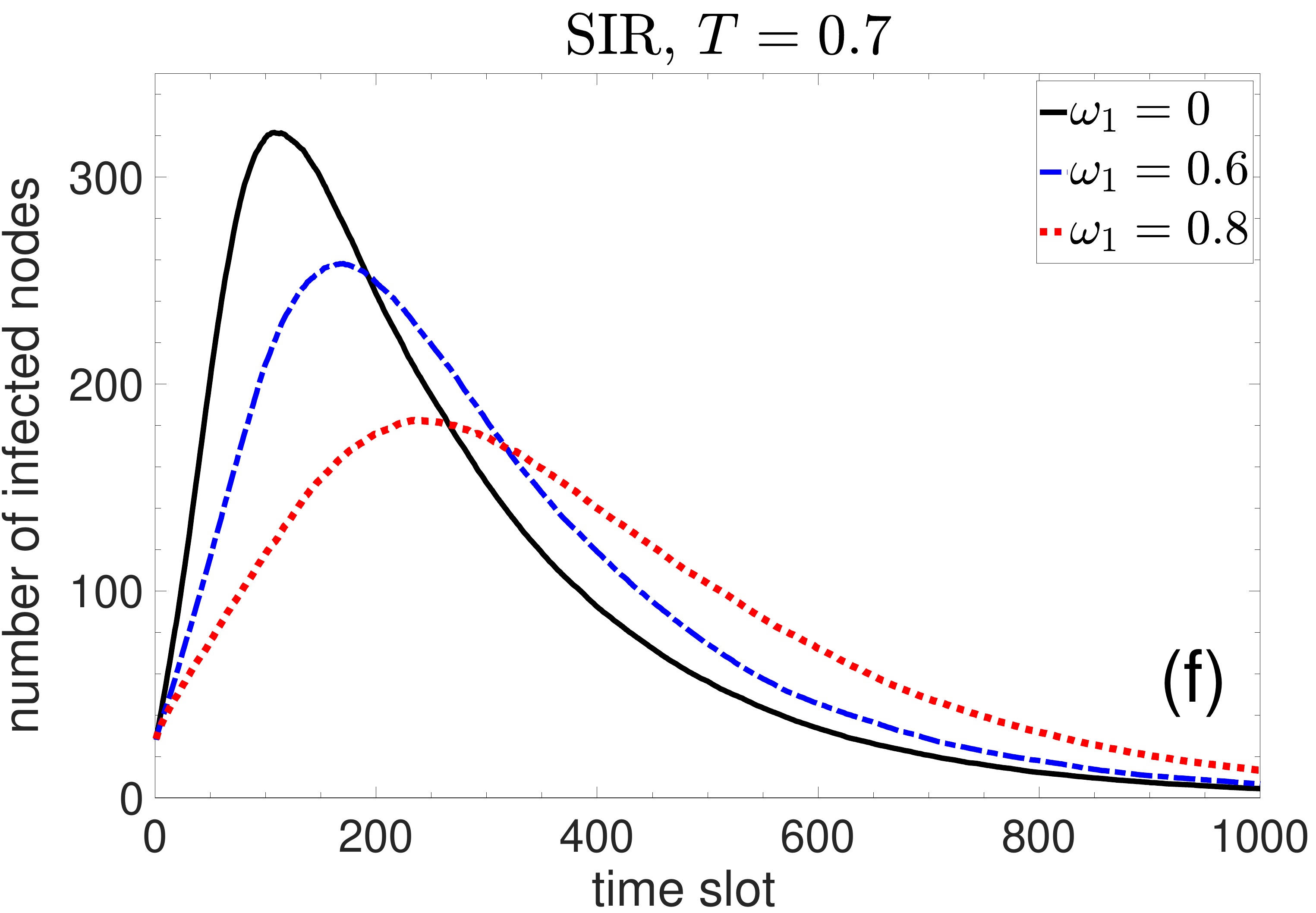}
\caption{Exploring epidemic spreading dynamics on temporal networks generated by the $(\omega_1, \omega_2)$-dynamic-$\mathbb{S}^1$ model. Plots (a)-(c) depict the number of infected nodes over time averaged over 100 simulations of the Susceptible-Infected-Susceptible (SIS) model~\cite{sis_ref}. There are $N=500$ nodes with a low expected degree $\kappa=\bar{k}=0.3$, yielding network snapshots in the disconnected regime, as in human proximity networks~\cite{flores2018,Papadopoulos2019}. Moving from left to right, the network temperature is $0.2$, $0.5$, and $0.7$. Results are presented for different levels of the link persistence probability $\omega_1$, while in all cases $\omega_2=0$. The total number of time slots is $\tau=1000$. The SIS model simulations start with $5\%$ of the nodes randomly infected, and the infection and recovery probabilities per slot are $0.5$ and $0.005$, respectively. Plots (d)-(f) show similar dynamics for the Susceptible-Infected-Recovered (SIR) model~\cite{sis_ref}, with all parameters the same as in (a)-(c). See Appendix~\ref{sec:appendix0} for further details and results illustrating the effect of non-link persistence.
\label{fig:epidemics}}
\end{figure*}

In addition to advancing modeling, incorporating distinct persistence probabilities for connections and disconnections, and understanding their effects on temporal network properties, is important for better understanding the behavior of dynamical processes running on temporal networks. This point is illustrated in Fig.~\ref{fig:epidemics} in the context of epidemic spreading. The figure shows that stronger link persistence can slow down spreading, depending on the setting. This occurs because pairs of nodes remain connected for longer durations, effectively reducing their opportunities to connect with and infect other nodes. Non-link persistence has a lesser effect unless it is very strong, see Appendix~\ref{sec:appendix0}. At the same time, Fig.~\ref{fig:epidemics} shows that spreading is also affected by the network temperature $T$, with lower values of $T$ suppressing spreading. This is because lower values of $T$ favor the localization of connections in the snapshots, as explained in Sec.~\ref{sec:S1}. We note that temporality has major implications not only within the context of epidemic spreading but also in many other contexts, such as wireless communications~\cite{ContiOppurtunisticOverview}, synchronization and diffusion~\cite{Masuda2013}, the evolution of cooperation~\cite{Li2020}, the emergence of chaos~\cite{Rock2023}, and the controllability of temporal networks~\cite{Li2017}. The ($\omega_1, \omega_2$)-dynamic-$\mathbb{S}^1$ is versatile and can be utilized in any context.

Despite the increased complexity introduced by the ($\omega_1, \omega_2$)-dynamic-$\mathbb{S}^1$, we can still analyze key properties of the model, including its connection probability function, the distributions of contact and intercontact durations, as well as the expected time-aggregated degree, elucidating their dependence on $\omega_1$, $\omega_2$, and the network temperature $T$. We focus on the (inter)contact distributions as they constitute perhaps the most fundamental characteristics affecting the performance of processes running on temporal networks~\cite{ContiOppurtunisticOverview, vazquez2007, timo2009, Karsai2011, machens2013, gauvin2013}. 
We show that the persistence probabilities $\omega_1$ and $\omega_2$ affect only the averages of these distributions but not their tails. Their tails follow power laws with exponents that depend only on the network temperature $T$, and these exponents are the same as in the case of $\omega_1=\omega_2$~\cite{Zambirinis2022}. Our results are proven for sufficiently large networks.

The expected time-aggregated degree represents the average number of distinct nodes that a node connects to during an observation period, and is another important characteristic of a temporal network~\cite{flores2018, StarniniDevices2017}. We show that as  $\omega_1$ or $\omega_2$ increases, or as $T$ decreases, the expected time-aggregated degree decreases, which can slow down dynamical processes such as spreading (Fig.~\ref{fig:epidemics}).
Having three independent parameters---$\omega_1$, $\omega_2$, and $T$---we can more flexibly adjust the average contact and intercontact durations, as well as the expected time-aggregated degree in the model. This finer control allows for a more nuanced exploration of temporal network dynamics and their impact on dynamical processes.

The rest of the paper is structured as follows. In the next section, we provide a brief overview of the $\mathbb{S}^{1}$ model. In Sec.~\ref{sec:omega_1_2_dynamic_S1}, we present the $(\omega_1, \omega_2)$-dynamic-$\mathbb{S}^1$ model and analyze its connection probability function. In Secs.~\ref{sec:contact} and~\ref{sec:intercontact}, we analyze the contact and intercontact distributions in the model, show their duality, and prove their power law tails. In Sec.~\ref{sec:degree}, we analyze the expected time-aggregated degree. In Sec.~\ref{sec:other_work}, we discuss the model in the context of other related work. Finally, in Sec.~\ref{sec:conclusion}, we outline open problems and interesting directions for future work, and conclude the paper.

%%%%%%%%%%%%%%%%%%%%%%%%%%%%%%%%%%%%%%%%%%%%%%%%%%%%%%%%%%%%%%%%%%%%%%%%%%%%%%%%%%%%%%%%%%%%%%%%%%%%%%%%%%%%%%%%%%%%%%%%
\section{Preliminaries}
\label{sec:S1}

In the $\mathbb{S}^{1}$ model~\cite{Krioukov2010}, each node is associated with a pair of hidden (or latent) variables $(\kappa, \theta)$. The hidden variable $\kappa$ represents the \emph{popularity} of the node, and is proportional to the node's expected degree in the network. The hidden variable $\theta$ represents the angular \emph{similarity} coordinate of the node on a circle of radius $R=N/2\pi$, where $N$ is the total number of nodes~\cite{Papadopoulos2012}. 

To generate a network that has size $N$, average node degree $\bar{k}$, and temperature $T \in (0,1)$, we perform the following steps:
\begin{enumerate}
\item[(i)] For each node $i \in \{1, 2,\ldots,N\}$, we sample its degree variable $\kappa_i$ from a probability density function (PDF) $\rho(\kappa)$, and its angular coordinate $\theta_i$ uniformly at random from $[0, 2\pi]$.
\item[(ii)] We connect every pair of nodes $i$ and $j$ according to the Fermi-Dirac connection probability
\begin{equation}
\label{eq:p_s1}
p_{ij} = \frac{1}{1+\chi_{ij}^{1/T}},
\end{equation}
\end{enumerate}
where $\chi_{ij}$ is the effective distance between nodes $i$ and $j$,
\begin{equation}
\label{eq:chi}
\chi_{ij} = \frac{R \Delta\theta_{ij}}{\mu \kappa_i \kappa_j}.
\end{equation}
In the above relation, $\Delta \theta_{ij}=\pi - | \pi -|\theta_i - \theta_j||$ represents the similarity distance between nodes $i$ and $j$. $\Delta \theta$ follows a uniform distribution on $[0, \pi]$, i.e., its PDF is $f(\Delta \theta)=1/\pi$. 

We are interested in sparse networks, where $N \gg \bar{k}$. In such cases, the resulting degree distribution in the network has a similar form as $\rho(\kappa)$~\cite{Boguna2003}. We also note that smaller values of the temperature $T$ favor connections at smaller effective distances, i.e., the localization of connections, increasing clustering in the network. Finally, parameter $\mu$ in Eq.~(\ref{eq:chi}) is derived from the requirement that the expected degree in the network is $\bar{k}$, yielding
\begin{equation}
\label{eq:mu}
\mu=\frac{\bar{k}\sin{(T \pi)}}{2\bar{\kappa}^2 T \pi},
\end{equation}
where $\bar{\kappa} = \int \kappa \rho(\kappa) \mathrm{d} \kappa$.

The $\mathbb{S}^{1}$ model is isomorphic to random hyperbolic graphs (RHGs) after a transformation of the degree variables $\kappa$ to radial coordinates $r$ on the hyperbolic disk (see Ref.~\cite{Krioukov2010} for more details).  

%%%%%%%%%%%%%%%%%%%%%%%%%%%%%%%%%%%%%%%%%%%%%%%%%%%%%%%%%%%%%%%%%%%%%%%%%%%%%%%%%%%%%%%%%%%%%%%%%%%%%%%%%%%%%%%%%%%%%%
\section{$(\omega_1, \omega_2)$-dynamic-$\mathbb{S}^{1}$}
\label{sec:omega_1_2_dynamic_S1}

The $(\omega_1,\omega_2)$-dynamic-$\mathbb{S}^{1}$ model generates a series of network snapshots, $G_t$, $t=1,\ldots, \tau$, where $\tau$ represents the total number of time slots. In the model, there are $N$ nodes that are assigned hidden variables $(\kappa, \theta)$ as in the $\mathbb{S}^{1}$ model, which remain fixed throughout the snapshots. The temperature $T \in (0,1)$ and the persistence probabilities $\omega_1 \in [0,1)$ and $\omega_2 \in [0,1)$ are also fixed. While each snapshot $G_t$ can potentially have a different average degree $\bar{k}_t$, to facilitate the analysis, we assume here a uniform average degree, i.e., $\bar{k}_t=\bar{k}$,~$\forall t$. Therefore, the model parameters are $N, \tau, \rho(\kappa), \bar{k}, T, \omega_1, \omega_2$. 

Let
\[
e_{ij}^{(t)} = \begin{cases}
	1  &  \text{if nodes~} i \text{~and~} j \text{~are connected at time } t, \\
	0  &  \text{otherwise}. 
\end{cases}
\]
The snapshots in the model are generated according to the following rules:
\begin{enumerate}
\item[(1)] Snapshot $G_1$ is generated according to the $\mathbb{S}^{1}$ model.
\item[(2)] At each time step $t=2, \ldots, \tau$, snapshot $G_t$ starts with $N$ disconnected nodes.
\item [(3)] Each pair of nodes $i, j$ in snapshot $G_t$ connects according to the following conditional connection probabilities:
\begin{align}
\label{eq:cp1b}
\mathbb{P}[e_{ij}^{(t)}=1|e_{ij}^{(t-1)}=1]&=\omega_1 + (1-\omega_1)\tilde{p}_{ij},\\
\label{eq:cp2b}
\mathbb{P}[e_{ij}^{(t)}=1|e_{ij}^{(t-1)}=0]&=(1-\omega_2)\tilde{p}_{ij},
\end{align}
where
\begin{equation}
\label{eq:p_s1omega}
\tilde{p}_{ij}=\frac{1}{1+\Big(\frac{1-\omega_2}{1-\omega_1}\Big)\chi_{ij}^{1/T}}.
\end{equation}
\item[(4)] At time $t+1$, the process is repeated to generate snapshot $G_{t+1}$.
\end{enumerate}
Equation~(\ref{eq:cp1b}) represents the scenario in which the pair $i, j$ is connected in the previous time slot $t-1$. In this case, the pair remains connected in slot $t$ either because the connection persists from $t-1$ (with probability $\omega_1$) or because the connection is established according to the probability $\tilde{p}_{ij}$. Equation~(\ref{eq:cp2b}) represents the situation where the pair $i, j$ is disconnected in $t-1$.  In this case, the pair can establish a connection in slot $t$ if the disconnection does not persist from $t-1$ (with probability $1-\omega_2$) and the connection is established according to $\tilde{p}_{ij}$. 

We note that $\omega_1$ has a greater influence on the stability of connections at larger effective distances, which would otherwise be of short duration. On the other hand, $\omega_2$ has a greater influence on the stability of disconnections at smaller effective distances that would otherwise be of short duration. Further, we note that a time slot in the model represents a discrete time step, which can correspond to any real-time duration depending on the scenario being modeled. For example, it can represent seconds or minutes in the case of human contact networks~\cite{Papadopoulos2019}, or days, weeks, or other durations in the case of other types of evolving networks~\cite{papaefthymiou2024}.

As we show below, the choice of the connection probability function in Eq.~(\ref{eq:p_s1omega}) ensures that the unconditional connection probability in the model is given by Eq.~(\ref{eq:p_s1}). Consequently, snapshots generated by the model are equivalent to RHGs, despite the dependencies introduced among them by the persistence probabilities $\omega_1$ and $\omega_2$.

%%%%%%%%%%%%%%%%%%%%%%%%%%%%%%%%%%%%%%%%%%%%%%%%%%%%%%%%%%%%%%%%%%%%%%%%%%%%%%%%%%%%%%%%%%%%%%%%%%%%%%%%%%%%%%%%%%%%%%
\emph{Unconditional connection probability}. We can express the unconditional connection probability for any node pair $i, j$ at time $t = 2, 3, \ldots$, as follows:
\begin{align}
\nonumber \mathbb{P}[e_{ij}^{(t)}=1] &= \mathbb{P}[e_{ij}^{(t)}=1|e_{ij}^{(t-1)}=1] \times \mathbb{P}[e_{ij}^{(t-1)}=1]\\ \nonumber &+ \mathbb{P}[e_{ij}^{(t)}=1|e_{ij}^{(t-1)}=0] \times(1-\mathbb{P}[e_{ij}^{(t-1)}=1])\\
\nonumber &= [\omega_1+(\omega_2-\omega_1)\tilde{p}_{ij}]\times \mathbb{P}[e_{ij}^{(t-1)}=1]\\ 
&+(1-\omega_2)\tilde{p}_{ij}.
\label{eq:prec}
\end{align}
Solving the above recurrence relation for $\mathbb{P}[e_{ij}^{(t)}=1]$, with the initial condition $\mathbb{P}[e_{ij}^{(1)}=1]=p_{ij}$, yields
\begin{equation}
\label{eq:recsol}
\mathbb{P}[e_{ij}^{(t)}=1]=\frac{B}{1-A}-A^{t-1}\left(\frac{B}{1-A}-p_{ij}\right),
\end{equation}
where $A=\omega_1+(\omega_2-\omega_1)\tilde{p}_{ij}$, and $B=(1-\omega_2)\tilde{p}_{ij}$. 

We observe that
\begin{align}
\frac{B}{1-A} =\frac{1}{1+\chi_{ij}^{1/T}}=p_{ij}.
\label{eq:obs}
\end{align}
Therefore, Eq.~(\ref{eq:recsol}) yields
\begin{equation}
\label{eq:equil}
\mathbb{P}[e_{ij}^{(t)}=1]=p_{ij},~\forall t.
\end{equation}
Thus, the unconditional connection probability is indeed as in Eq.~(\ref{eq:p_s1}). In the next section, we analyze the distribution of contact durations in the model.

%%%%%%%%%%%%%%%%%%%%%%%%%%%%%%%%%%%%%%%%%%%%%%%%%%%%%%%%%%%%%%%%%%%%%%%%%%%%%%%%%%%%%%%%%%%%%%%%%%%%%%%%%%%%%%%%%%%%%%%%
\section{Distribution of contact durations}
\label{sec:contact}

Let $\tau$ be the total number of time slots during which we observe the system. To derive the contact distribution, we need to consider the probability of observing a sequence of exactly $t$ consecutive time slots where two nodes $i$ and $j$ with hidden degrees $\kappa_i$ and $\kappa_j$ and angular distance $\Delta \theta_{ij}$ are connected. Any such sequence should be enclosed within two slots where the two nodes are not connected. That is, we ignore for now the boundary cases where the first or last of the $t$ slots starts or ends at the beginning or end of the observation period $\tau$. Therefore, $t$ ranges from $1$ to $\tau-2$. We denote this probability by $r_\textnormal{c}(t; \kappa_i, \kappa_j, \Delta\theta_{ij})$.

We note that given a sequence of length $t$, there exist $\tau-t-1$ possible starting positions for this sequence. For example, if $t=3$, the nodes can be disconnected in slot $s-1$, connected in slots $s, s+1, s+2$, and disconnected in slot $s+3$, where $s$ ranges from $2$ to $\tau-3$. Consequently, the probability of observing a slot where a sequence of length $t$ can start is
\begin{equation}
\label{eq:g_tau}
g_\tau(t)=\frac{\tau-t-1}{\tau}.
\end{equation}
Furthermore, we observe the following:
\begin{enumerate}
\item[(i)] The unconditional probability that two nodes $i$ and $j$ are disconnected in a slot $s$ is $1-p_{ij}$, where $p_{ij}$ is given by Eq.~(\ref{eq:p_s1}).
\item[(ii)] Given that they are disconnected in slot $s$, the probability that $i$ and $j$ are connected in slot $s+1$ is $(1-\omega_2) \tilde{p}_{ij}$, where $\tilde{p}_{ij}$ is given by Eq.~(\ref{eq:p_s1omega}).
\item[(iii)] Given that they are connected in slot $s+1$, the probability that $i$ and $j$ remain connected in slots $s+2, \ldots, s+t$ is $[\omega_1 + (1-\omega_1)\tilde{p}_{ij} ]^{t-1}$.
\item[(iv)] Finally, given that they are connected in slot $s+t$, the probability that $i$ and $j$ are disconnected in slot $s+t+1$ is $(1-\omega_1) (1-\tilde{p}_{ij})$.
\end{enumerate}
The probability $r_\textnormal{c}(t;\kappa_i,\kappa_j,\Delta\theta_{ij})$ is obtained by multiplying $g_\tau(t)$ with the probabilities described in points (i) to (iv) above,
\begin{widetext}
\begin{equation}
\label{eq:p_c}
r_\textnormal{c}(t;\kappa_i,\kappa_j,\Delta\theta_{ij})=g_\tau(t)(1-\omega_1)(1-\omega_2)(1-p_{ij})\tilde{p}_{ij}(1-\tilde{p}_{ij})[\omega_1+(1-\omega_1)\tilde{p}_{ij}]^{t-1}. 
\end{equation}
\end{widetext}

The contact distribution, denoted as $P_{\textnormal{c}}(t)$ and defined for $t\geq 1$, is given by
\begin{equation}
\label{eq:p_c_norm}
P_{\textnormal{c}}(t) = \frac{r_\textnormal{c}(t)}{\sum_{j} r_\textnormal{c}(j)} \propto r_\textnormal{c}(t).
\end{equation}
In the last expression, $r_\textnormal{c}(t)$ is determined by removing the conditions on $\kappa_i$, $\kappa_j$, and $\Delta \theta_{ij}$ from Eq.~(\ref{eq:p_c}),
\begin{widetext}
\begin{equation}
r_\textnormal{c}(t) = \int \int \int r_\textnormal{c}(t ; \kappa, \kappa', \Delta\theta) \rho(\kappa) \rho(\kappa') f(\Delta \theta) \mathrm{d} \kappa \mathrm{d} \kappa' \mathrm{d} \Delta\theta.
\end{equation}
\end{widetext}
We note that in practice, given a set of nonzero contact durations, the empirical $P_\textnormal{c}(t)$ is determined by the ratio $n_t/\sum_j n_j$, where $n_t$ represents the number of contact durations in the set with length $t$.

Removing the condition on $\Delta \theta_{ij}$ from Eq.~(\ref{eq:p_c}), yields
\begin{widetext}
\begin{align} 
\nonumber r_\textnormal{c}(t ; \kappa_i, \kappa_j)&=\frac{1}{\pi} \int_0^\pi r_\textnormal{c}(t;\kappa_i, \kappa_j, \Delta\theta) \mathrm{d} \Delta\theta\\
\nonumber &= g_\tau(t) \frac {2 \mu \kappa_i \kappa_j T}{N} (1-\omega_1)^{1+T}(1-\omega_2)^{1-T}\omega_1^{t-1}  \int_{u_0^{ij}}^{1} u^{-T} (1-u)^{1+T} \Big(1-\frac{\omega_1-1}{\omega_1}u\Big)^{t-1}\Big(1-\frac{\omega_2-\omega_1}{1-\omega_1}u\Big)^{-1} \mathrm{d}u,\\
\label{eq:full_contact_integral}
&\textnormal{where~}
u_0^{ij} = \frac{1}{1 + \Big(\frac{1-\omega_2}{1-\omega_1}\Big)\Big(\frac{N}{2\mu\kappa_i \kappa_j}\Big)^{1/T}}.
\end{align}
\end{widetext}
To obtain the above relation, we performed the change of integration variable
$u = 1/[1 + (\frac{1-\omega_2}{1-\omega_1})(\frac{N \Delta \theta}{2 \pi \mu \kappa_i \kappa_j})^{1/T}]$.

Now, for sufficiently large network sizes $N$, $u_0^{ij}$ tends to zero. This allows us to remove the condition on $\kappa_i$ and $\kappa_j$ from Eq.~(\ref{eq:full_contact_integral}), and write, \emph{irrespective of the form of $\rho(\kappa)$},
\begin{widetext}
\begin{equation}
r_\textnormal{c}(t) \approx g_\tau(t) \frac {2 \mu \bar{\kappa}^2 T}{N} (1-\omega_1)^{1+T}(1-\omega_2)^{1-T}\omega_1^{t-1} \int_0^1 u^{-T} (1-u)^{1+T} \Big(1-\frac{\omega_1-1}{\omega_1}u \Big)^{t-1}
\Big(1-\frac{\omega_2-\omega_1}{1-\omega_1}u\Big)^{-1} \mathrm{d}u.
\label{eq:rc_approx1} 
\end{equation}
\end{widetext}

The integral in Eq.~(\ref{eq:rc_approx1}) can be evaluated numerically. However, we observe that it is in a form suitable for representation using the Appell $F_1$ series~\cite{bateman1953higher}. This representation will be employed below to deduce the behavior of the tail of $r_\textnormal{c}(t)$. In particular, Émile Picard discovered in 1881 that the Appell $F_1$ series, whose definition is provided in Appendix~\ref{sec:appendix1}, has the following Euler-type integral representation (cf.~section 5.8.2 of Ref.~\cite{bateman1953higher}):
\begin{widetext}
\begin{equation}
F_1[a, b_1, b_2, c; x, y] =\frac{\Gamma{(c)}}{\Gamma{(a)}\Gamma{(c-a)}}\int_0^1 u^{a-1} (1-u)^{c-a-1}(1-x u)^{-b_1} (1-y u)^{-b_2}\mathrm{d}u.
\label{eq:picard}
\end{equation}
\end{widetext}
The above relation is valid for $c > a > 0$, and $\Gamma$ is the gamma function. Utilizing this representation with $\alpha=1-T$, $b_1=1-t$, $b_2=1$, $c=3$, $x=\frac{\omega_1-1}{\omega_1}$, and $y=\frac{\omega_2-\omega_1}{1-\omega_1}$, substituting $\mu$ with its expression in Eq.~(\ref{eq:mu}), and employing the identity $\frac{\pi}{\sin{(T \pi)}}=\Gamma{(1-T)}\Gamma{(T)}$, we can rewrite Eq.~(\ref{eq:rc_approx1}), as
\begin{widetext}
\begin{align}
\nonumber r_\textnormal{c}(t) &\approx  g_\tau(t) \frac{\bar{k} T (1+T)}{2 N} (1-\omega_1)^{1+T}(1-\omega_2)^{1-T}\omega_1^{t-1} F_1[1-T, 1-t, 1, 3; \frac{\omega_1-1}{\omega_1}, \frac{\omega_2-\omega_1}{1-\omega_1}]\\
&=g_\tau(t) \frac{\bar{k} T (1+T)}{2 N} (1-\omega_1)^{2+T}(1-\omega_2)^{-T} F_1[2+T, 1-t, 1, 3; 1-\omega_1, \frac{\omega_1-\omega_2}{1-\omega_2}].
\label{eq:rc_appel}
\end{align}
\end{widetext}
The last equality is obtained by performing the change of variable $v=1-u$ in the integral of Eq.~(\ref{eq:picard}), or equivalently, by applying the transformation given by Eq.~(1) in section~5.11 of Ref.~\cite{bateman1953higher}. For $\omega_1=\omega_2=\omega$ the last $F_1$ function in Eq.~(\ref{eq:rc_appel}) degenerates to the Gauss hypergeometric function ${}_2 F_{1}[2+T, 1-t, 3; 1-\omega]$ (see Appendix~\ref{sec:appendix1} for its definition), and we recover the relation for $r_\textnormal{c}(t)$ found in Ref.~\cite{Zambirinis2022}.

%%%%%%%%%%%%%%%%%%%%%%%%%%%%%%%%%%%%%%%%%%%%%%%%%%%%%%%%%%%%%%%%%%%%%%%%%%%%%%%%%%%%%%%%%%%%%%%%%%%%%%%%%%%%%%%%%%%%%%%%
\emph{Boundary cases}. The preceding analysis did not consider the boundary case where the first slot in the sequence of $t$ slots, during which two nodes are connected, starts at the beginning of the observation period $\tau$. In this case,  $g_\tau(t)=1/\tau$, and the probability of observing this event for two nodes $i$ and $j$ is given by
\begin{widetext}
\begin{equation}
\label{eq:p_c_boundary_1}
r_\textnormal{c}^{\textnormal{b}}(t;\kappa_i,\kappa_j,\Delta\theta_{ij})=\frac{1}{\tau}(1-\omega_1)p_{ij}(1-\tilde{p}_{ij})[\omega_1 + (1-\omega_1)\tilde{p}_{ij} ]^{t-1},
\end{equation}
\end{widetext}
for $t=1, \ldots, \tau-1$. Similarly, the analysis did not consider the case where the last slot in the sequence of $t$ slots, during which two nodes are connected, finishes at the end of the observation period. It is easy to see that the probability of observing this event is also given by Eq.~(\ref{eq:p_c_boundary_1}).

Following the same procedure to remove the conditions on $\kappa_i$, $\kappa_j$, and $\Delta \theta_{ij}$, and employing the same transformations as before, we can write that the total probability for these two cases is given by
\begin{widetext}
\begin{align}
\nonumber r_\textnormal{c}^{\textnormal{b}}(t) &\approx \frac{2}{\tau} \frac {2 \mu \bar{\kappa}^2 T}{N} (1-\omega_1)^{T}(1-\omega_2)^{1-T}\omega_1^{t-1} \int_0^1 u^{-T} (1-u)^T \Big(1-\frac{\omega_1-1}{\omega_1}u \Big)^{t-1}
\Big(1-\frac{\omega_2-\omega_1}{1-\omega_1}u\Big)^{-1} \mathrm{d}u\\
&= \frac{2}{\tau} \frac {\bar{k} T}{N} (1-\omega_1)^{1+T}(1-\omega_2)^{-T} F_1[1+T, 1-t, 1, 2; 1-\omega_1, \frac{\omega_1-\omega_2}{1-\omega_2}].
\label{eq:rc_approx1boundary} 
\end{align}
\end{widetext}
We note that for any finite $t$, $r_\textnormal{c}^{\textnormal{b}}(t)$ tends to zero as $\tau \to \infty$. However, as $t$ approaches $\tau$, the contribution of these boundary cases becomes significant. Accounting for these cases, the combined probability of observing a sequence of $t$ consecutive slots in which two nodes are connected is given by
\begin{equation}
\label{eq:combined}
\tilde{r}_\textnormal{c}(t)= r_\textnormal{c}(t)+r_\textnormal{c}^{\textnormal{b}}(t),
\end{equation}
for $t=1, \ldots, \tau-1$.

\begin{figure*}
\includegraphics[width=3in]{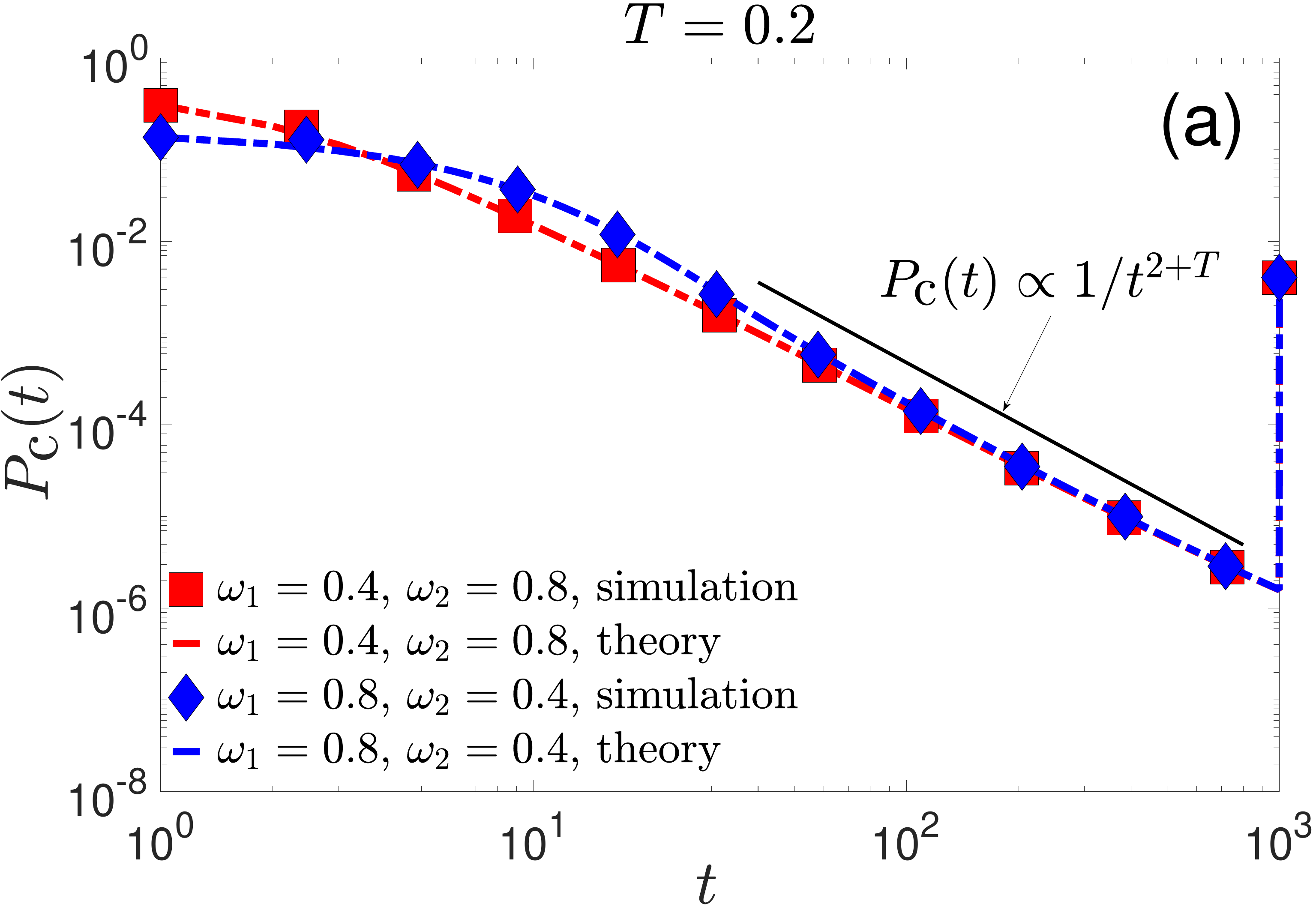}
\includegraphics[width=3in]{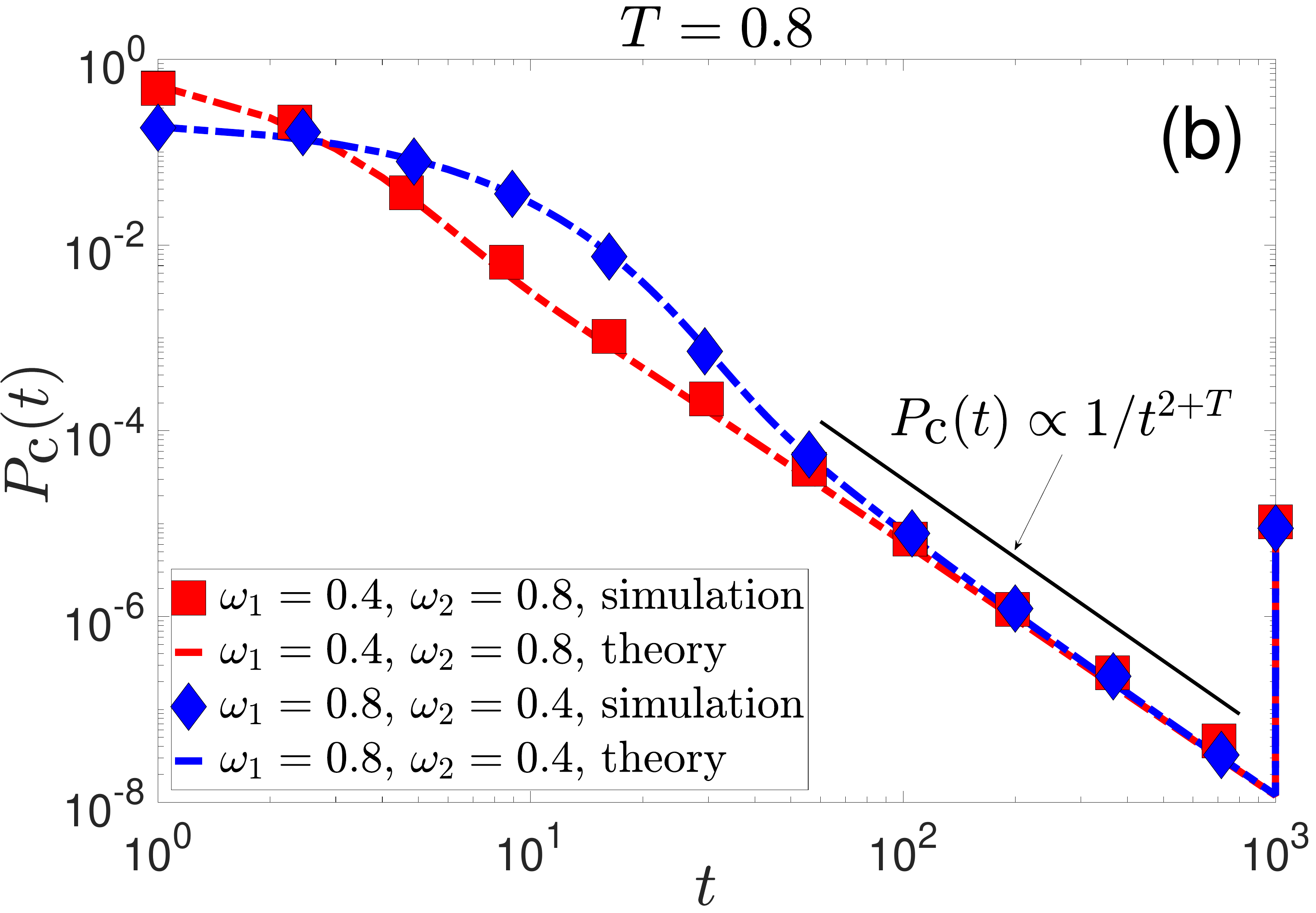}
\caption{Distribution of contact durations in simulated networks with the $(\omega_1, \omega_2)$-dynamic-$\mathbb{S}^{1}$ model~vs.~theoretical predictions. The latter are given by~$P_{\textnormal{c}}(t)=\tilde{r}_\textnormal{c}(t)/\sum_{j=1}^\tau\tilde{r}_\textnormal{c}(j)$, where $\tilde{r}_\textnormal{c}(t)$ is given by Eq.~(\ref{eq:combined}) for $t=1, \ldots, \tau-1$, and by Eq.~(\ref{eq:rc_approx2boundary}) for $t=\tau$ (yielding the rightmost point on the plots). The number of nodes is $N=500$, the average node degree is $\bar{k}=5$, all nodes have the same expected degree $\kappa=\bar{k}$, and the total number of time slots is $\tau=1000$. The network temperature in (a) is $T=0.2$, and in (b) $T=0.8$. Results are presented for two combinations of the persistence probabilities $\omega_1$ and $\omega_2$. The simulations are averaged over $10$ runs, and empirical distributions are logarithmically binned, excluding the rightmost point. Theoretical predictions are represented by dashed lines. Solid black lines show the power-law scaling  $P_{\textnormal{c}}(t) \propto 1/t^{2+T}$, deduced by Eq.~(\ref{eq:psscaling}). All axes use a logarithmic scale. 
\label{fig:val1}}
\end{figure*}

The final boundary case occurs when two nodes $i$ and $j$ remain connected for the entire observation period $\tau$. The probability of observing this case is
\begin{equation}
\label{eq:p_c_boundary_2}
r_\textnormal{c}^{\textnormal{b}}(\tau; \kappa_i,\kappa_j,\Delta\theta_{ij})=\frac{1}{\tau}p_{ij} [\omega_1 + (1-\omega_1)\tilde{p}_{ij} ]^{\tau-1}. 
\end{equation}
Removing the conditions on $\kappa_i$, $\kappa_j$, and $\Delta \theta_{ij}$, and employing the same transformations as before, gives
\begin{widetext}
\begin{align}
\nonumber r_\textnormal{c}^{\textnormal{b}}(\tau) &\approx \frac{1}{\tau} \frac {2\mu\bar{\kappa}^2T}{N}\Big(\frac{1-\omega_2}{1-\omega_1}\Big)^{1-T}\omega_1^{\tau-1} \int_0^1 u^{-T} (1-u)^{T-1} \Big(1-\frac{\omega_1-1}{\omega_1}u \Big)^{\tau-1}
\Big(1-\frac{\omega_2-\omega_1}{1-\omega_1}u\Big)^{-1} \mathrm{d}u\\
&= \frac{1}{\tau} \frac {\bar{k}}{N}\Big(\frac{1-\omega_1}{1-\omega_2}\Big)^T F_1[T, 1-\tau, 1, 1; 1-\omega_1, \frac{\omega_1-\omega_2}{1-\omega_2}].
\label{eq:rc_approx2boundary} 
\end{align}
\end{widetext}
We note that previous studies related to the dynamic-$\mathbb{S}^{1}$ model~\cite{Papadopoulos2019, Zambirinis2022} have not considered the above boundary cases. In Fig.~\ref{fig:val1}, we validate the above analysis with simulations, while also taking into account the boundary cases. In all cases, we calculate $r_{\textnormal{c}}(t)$ and $r_\textnormal{c}^{\textnormal{b}}(t)$ using their integral representations, as we have found it more efficient than utilizing the corresponding Appell $F_1$ series.

\begin{figure*}
\includegraphics[width=2.9in]{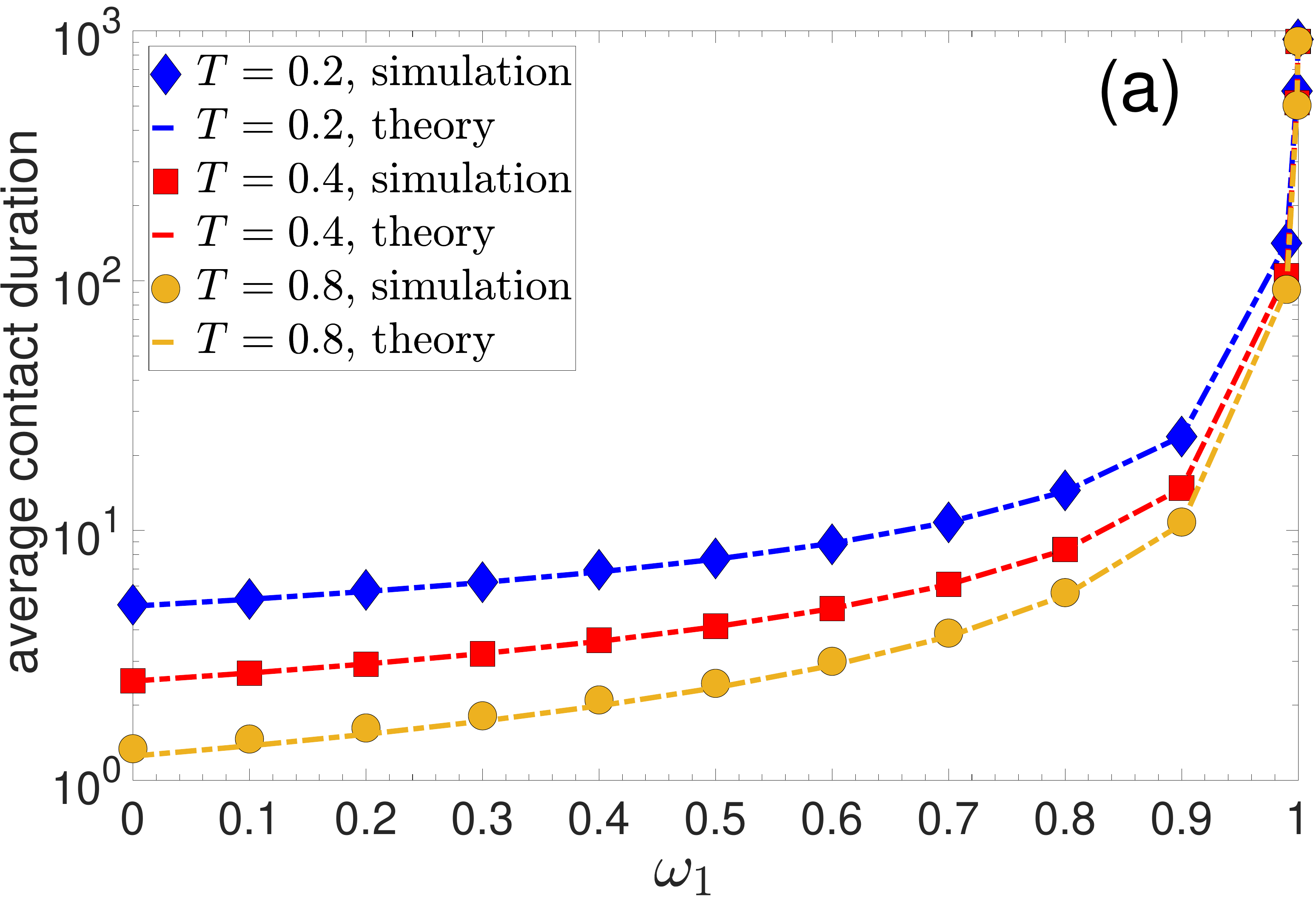}
\includegraphics[width=2.9in]{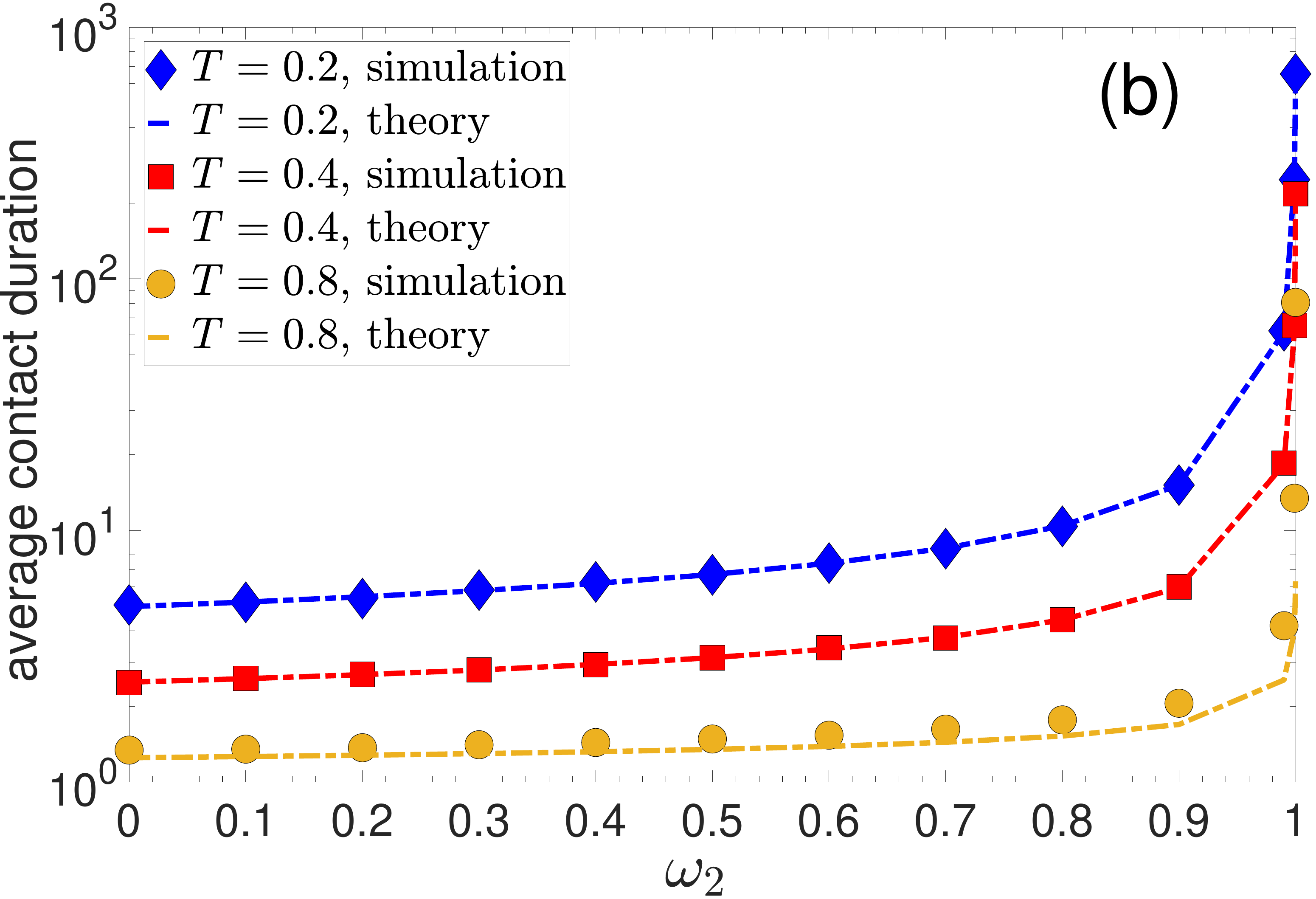}
\caption{Average contact duration vs. $\omega_1$, $\omega_2$, and $T$. Plot (a) shows the average contact duration in time slots as a function of the persistence probability of connections $\omega_1$. The persistence probability of disconnections, $\omega_2$, is set to zero. Results are shown for different values of the network temperature $T$. In each case the three rightmost points correspond respectively to $\omega_1 = 0.99$, $0.999$, and $0.9999$. All other parameters are the same as in Fig.~\ref{fig:val1}. The dashed lines depict theoretical predictions given by $\bar{t}_\textnormal{c}=\sum_{t=1}^{\tau} t P_\textnormal{c}(t)$, where $P_\textnormal{c}(t)$ is computed as in Fig.~\ref{fig:val1}. Plot (b) is similar to (a), except that $\omega_1$ is set to zero, and we vary $\omega_2$. The $y$-axes use a logarithmic scale. Deviations of analytical predictions from simulation results are due to finite network size effects and are more pronounced for values of $T$ or $\omega_2$ closer to 1.
\label{fig:ave_contact}}
\end{figure*}

%%%%%%%%%%%%%%%%%%%%%%%%%%%%%%%%%%%%%%%%%%%%%%%%%%%%%%%%%%%%%%%%%%%%%%%%%%%%%%%%%%%%%%%%%%%%%%%%%%%%%%%%%%%%%%%%%%%%%%%%
\emph{Average contact duration.} It is evident from our analysis and Fig.~\ref{fig:val1} that all three parameters—$\omega_1$, $\omega_2$, and $T$—affect the contact distribution. In Fig.~\ref{fig:ave_contact}, we investigate how these parameters affect the average contact duration. 

We see from Fig.~\ref{fig:ave_contact} that the average contact duration increases as either $\omega_1$ or $\omega_2$ increases, with the rate of increase becoming more pronounced as these parameters approach $1$. Moreover, we observe that the average contact duration is more sensitive to and increases more rapidly with $\omega_1$ than with $\omega_2$. This is expected, as $\omega_1$ directly impacts the probability that two nodes remain connected, given by Eq.~(\ref{eq:cp1b}). In particular, as $\omega_1 \to 1$, the probability in Eq.~(\ref{eq:cp1b}) approaches $1$, irrespective of the value of $\omega_2$. On the other hand, $\omega_2$ indirectly affects this probability via $\tilde{p}_{ij}$ (Eq.~(\ref{eq:p_s1omega})). Indeed, as $\omega_2 \to 1$, $\tilde{p}_{ij} \to 1$, and Eq.~(\ref{eq:cp1b}) tends to $1$, irrespective of the value of $\omega_1$. In other words, as $\omega_1 \to 1$ or $\omega_2 \to 1$, the contact distribution degenerates to $P_{\textnormal{c}}(t) \to 1$ for $t=\tau$, and $P_{\textnormal{c}}(t) \to 0$, for $t < \tau$, while the average contact duration tends to the value of the observation interval $\tau$. This convergence occurs faster with $\omega_1 \to 1$ than with $\omega_2 \to 1$.

Lastly, Fig.~\ref{fig:ave_contact} shows that the average contact duration also increases as $T$ decreases. A lower $T$ favors connections at smaller effective distances, thereby increasing the probability that connected pairs remain connected. For $T \to 0$, we obtain the same result as in the case of $\omega_1 \to 1$ or $\omega_2 \to 1$. 

%%%%%%%%%%%%%%%%%%%%%%%%%%%%%%%%%%%%%%%%%%%%%%%%%%%%%%%%%%%%%%%%%%%%%%%%%%%%%%%%%%%%%%%%%%%%%%%%%%%%%%%%%%%%%%%%%%%%%%%%
\emph{Tail of the contact distribution.} We conclude our analysis in this section by deducing the behavior of $P_{\textnormal{c}}(t)$ at large $t$. To this end, we utilize an asymptotic result given by Eq.~(20) in section 3.5.1 of Ref.~\cite{appellasymptotics2013}. This result states that for $x < 0$ and $|y| < 1$, we can express the Appell function $F_1 [a, b+\lambda, b', c; x, y]$ as a sum of Gauss hypergeometric functions,
\begin{widetext}
\begin{equation}
\label{eq:asymptres}
F_1 [a, b+\lambda, b', c; x, y]=\sum_{n=0}^{m-1}\binom{-b'}{n}\frac{(a)_n (-y)^n}{(c)_n} {_2}F_1 [b+\lambda, a+n, c+n; x] + O(\lambda^{-m-a}),
\end{equation} 
\end{widetext}
where $(q)_n$ denotes the Pochhammer symbol, defined as: $(q)_n=1$ for $n=0$ and $(q)_n=q(q+1) \ldots (q+n-1)$ for $n > 0$. Furthermore, we utilize the transformation given by Eq.~(2) in section~5.11 of Ref.~\cite{bateman1953higher}, which states that
\begin{widetext}
\begin{equation}
\label{eq:trans2}
F_1[a, b, b', c; x, y]=(1-x)^{-a}F_1 [a, c-b-b', b', c; \frac{x}{x-1}, \frac{y-x}{1-x}].
\end{equation}  
\end{widetext}
Using the above transformation, we can rewrite the $F_1$ function in Eq.~(\ref{eq:rc_appel}), which we refer to as $h_1$, as
\begin{widetext}
\begin{equation}
\label{eq:h_1}
h_1 \coloneqq F_1 [2+T, 1-t, 1,3; 1-\omega_1, \frac{\omega_1-\omega_2}{1-\omega_2}] = \omega_1^{-(2+T)} F_1 [2+T, 1+t, 1, 3; 1-\frac{1}{\omega_1}, 1-\frac{1-\omega_1}{\omega_1(1-\omega_2)}].
\end{equation}
\end{widetext}
Now, using Eq.~(\ref{eq:asymptres}) with $a=2+T$, $b=1$, $\lambda=t$, $b'=1$, $c=3$, $x=1-\frac{1}{\omega_1}$, and  $y=1-\frac{1-\omega_1}{\omega_1(1-\omega_2)}$, we can write
\begin{widetext}
\begin{equation}
\label{eq:h1expansion}
h_1=\omega_1^{-(2+T)}\sum_{n=0}^{m-1}\frac{(2+T)_n}{(3)_n}\Big( 1-\frac{1-\omega_1}{\omega_1(1-\omega_2)} \Big)^n {_2}F_1[1+t, 2+T+n, 3+n; 1-\frac{1}{\omega_1}] + O\Big(\frac{1}{t^{2+T+m}}\Big).
\end{equation} 
\end{widetext}
To write the above relation, we also utilized that $\binom{-1}{n}=(-1)^n$ for $n \in \mathbb{N}$. 

\begin{figure}[!]
\includegraphics[width=3in]{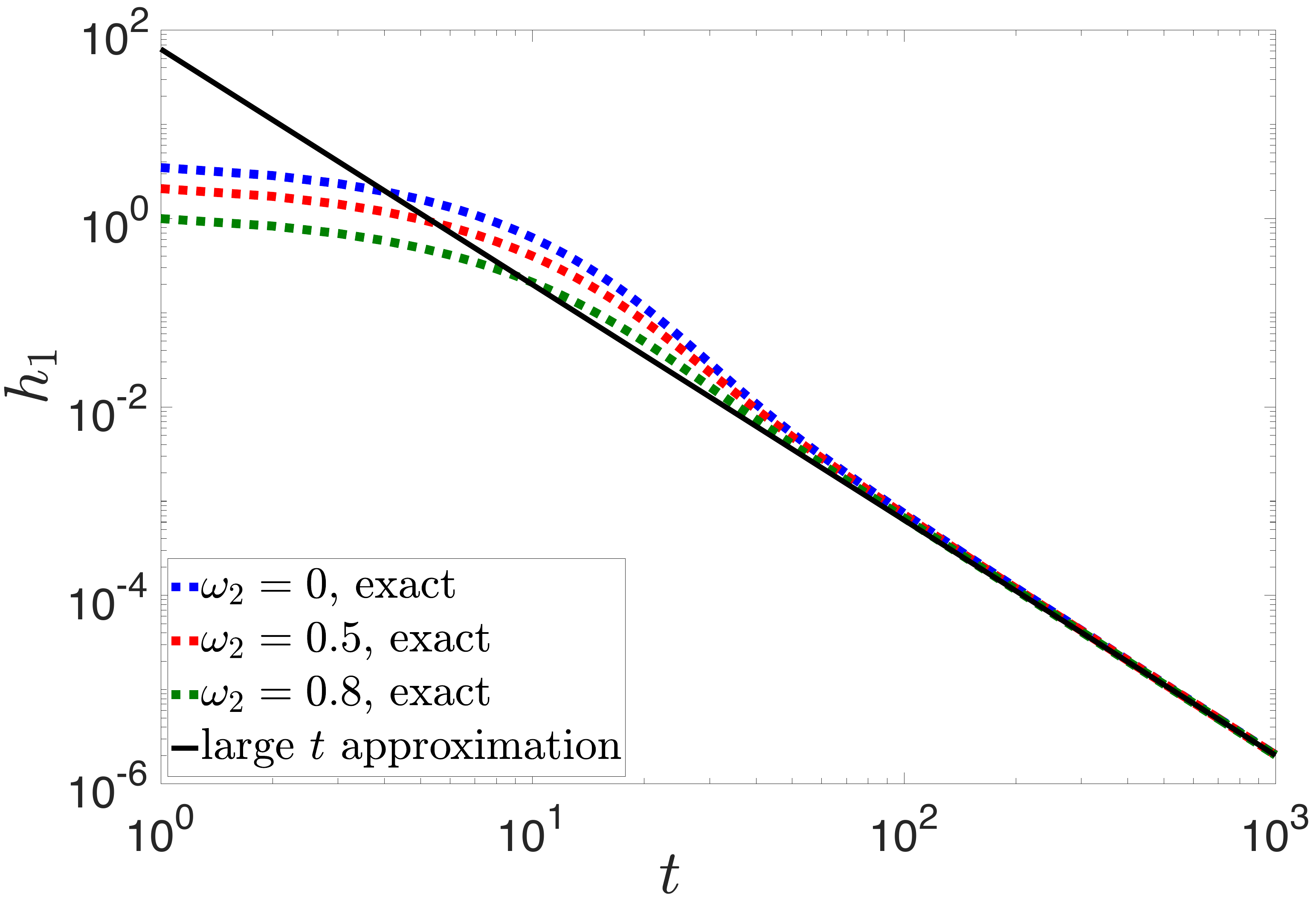}
\caption{
Function $h_1$ in Eq.~(\ref{eq:h_1}) (dotted lines) vs. the approximation for large $t$ in Eq.~(\ref{eq:h1approx}) (solid line). Results are shown for different values of $\omega_2$, while $\omega_1=0.8$ and $T=0.5$. All axes use a logarithmic scale.
\label{fig:approx1_val}}
\end{figure}

As shown in Appendix~\ref{sec:appendix2}, the ${_2}F_1$ function inside the sum in Eq.~(\ref{eq:h1expansion}) can be approximated for large $t$ as
\begin{widetext}
\begin{equation}
{_2}F_1[1+t, 2+T+n, 3+n; 1-\frac{1}{\omega_1}] \approx \frac{\Gamma{(3+n)(1/\omega_1-1)^{-(2+T+n)}}}{\Gamma{(1-T)}}\frac{1}{t^{2+T+n}}.
\label{eq:f1approx}
\end{equation}
\end{widetext}
Consequently, at large $t$, the term corresponding to $n=0$ in Eq.~(\ref{eq:h1expansion}) dominates, and we can approximate $h_1$ as

\begin{equation}
\label{eq:h1approx}
h_1 \approx \frac{2 (1-\omega_1)^{-(2+T)}}{\Gamma(1-T)}\frac{1}{t^{2+T}}. 
\end{equation} 
This approximation is validated in Fig.~\ref{fig:approx1_val}.

We note that Eq.~(\ref{eq:h1expansion}) holds for $x=1-\frac{1}{\omega_1} < 0$ and $|y|=\big| 1-\frac{1-\omega_1}{\omega_1(1-\omega_2)} \big| < 1$. The first inequality always holds (as $\omega_1 < 1$), while the second imposes the constraint $\omega_2 < \frac{3\omega_1-1}{2\omega_1}$. Additionally, the approximation in Eq.~(\ref{eq:f1approx}) requires $|1-\frac{1}{\omega_1}|<1$, which imposes the constraint $\omega_1 > 1/2$. Combined, these constraints define the region $\mathcal{R}_1$ of $\omega_1$ and $\omega_2$ depicted in Fig.~\ref{fig:region1}, for which the preceding analysis leading to~Eq.~(\ref{eq:h1approx}) holds. However, in Appendix~\ref{sec:appendix2}, we prove that Eq.~(\ref{eq:h1approx}), which is established here for the region $\mathcal{R}_1$, holds in fact true for any combination of $\omega_1, \omega_2 \in [0,1)$.
\begin{figure}[!]
\includegraphics[width=2in]{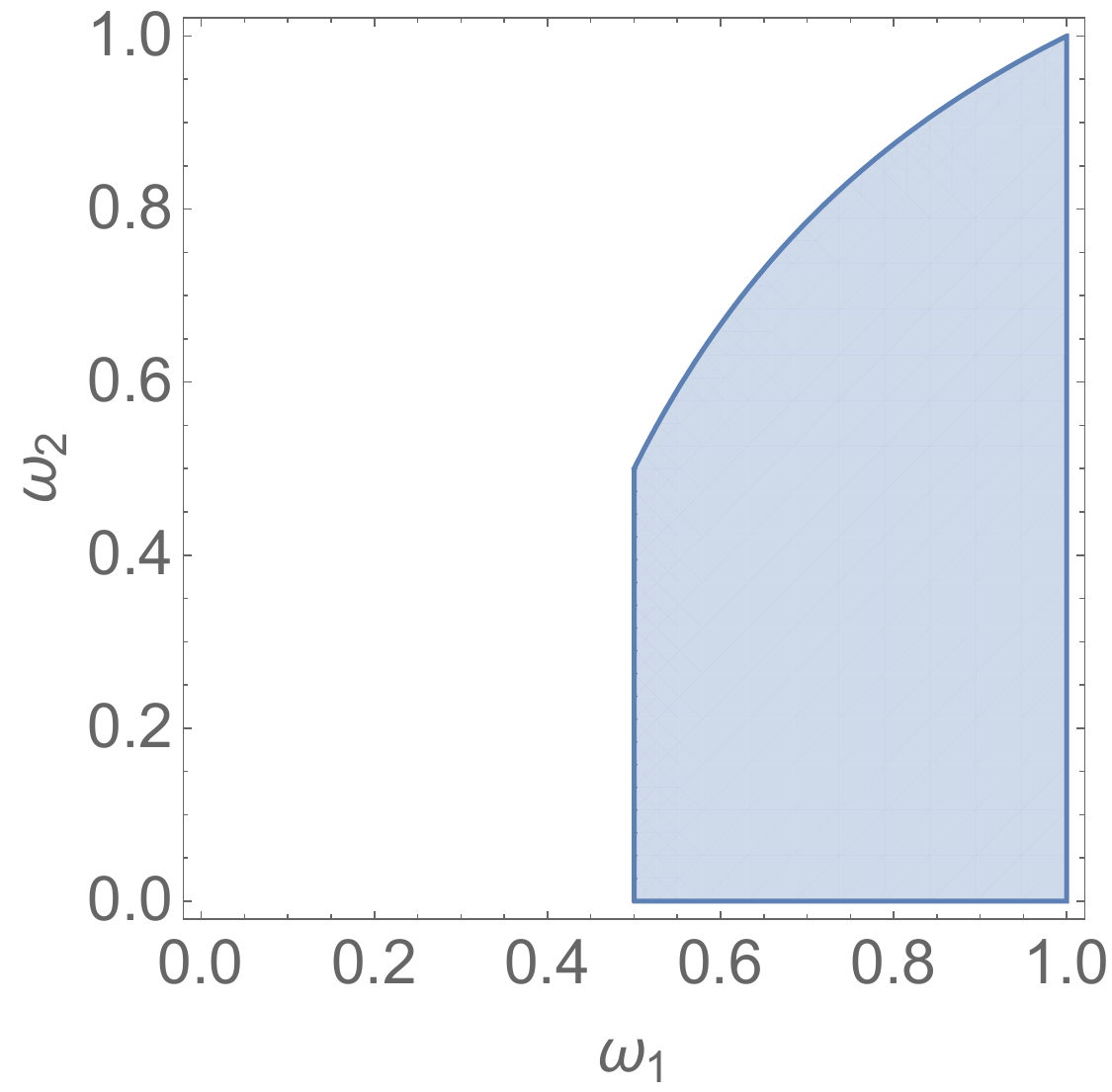}
\caption{Region $\mathcal{R}_1:=\{(\omega_1, \omega_2) \in \mathbb{R}^2 \ \mid \ \frac{1}{2}<\omega_1<1, \ 0 \leq \omega_2 < \frac{3\omega_1-1}{2\omega_1}\}$, shown as the blue-shaded area in the figure. In this region, Eqs.~(\ref{eq:h1expansion}) and~(\ref{eq:f1approx}) both hold, leading to Eq.~(\ref{eq:h1approx}).} 
\label{fig:region1}
\end{figure}

The above analysis (and the corresponding analysis in Appendix~\ref{sec:appendix2}) can be repeated for the function $F_1$ in Eq.~(\ref{eq:rc_approx1boundary}), which corresponds to the boundary cases. This yields, for large $t$,
\begin{align}
\nonumber F_1[1+T, 1-t, 1, 2; 1-\omega_1, \frac{\omega_1-\omega_2}{1-\omega_2}] \\
\approx \frac{(1-\omega_1)^{-(1+T)}}{\Gamma(1-T)}\frac{1}{t^{1+T}}.
\label{eq:h1approxb}
\end{align}

Utilizing the approximations given by Eqs.~(\ref{eq:h1approx}) and~(\ref{eq:h1approxb}), we can approximate $\tilde{r}_\textnormal{c}(t)$ in Eq.~(\ref{eq:combined}) for large $t$ as
\begin{align}
\nonumber \tilde{r}_\textnormal{c}(t) & \approx \frac{\bar{k} T}{N}\frac{(1-\omega_2)^{-T}}{\Gamma(1-T)}\Big[g_\tau(t)\frac{(1+T)}{t^{2+T}}+\frac{2}{\tau}\frac{1}{t^{1+T}}\Big]\\
&\propto \frac{1+T+(1-T)t/\tau}{t^{2+T}}.
\label{eq:psscaling}
\end{align}
The numerator in Eq.~(\ref{eq:psscaling}) is a sum of a constant ($1 + T$) and the linearly increasing term $(1-T)t/\tau$, which is upper-bounded by $1-T$. For $t \ll \tau$, such that $t/\tau \approx 0$, this term is insignificant. Therefore, $\tilde{r}_\textnormal{c}(t)$ and consequently the contact distribution $P_{\textnormal{c}}(t)$ decay according to the power law $1/t^{2+T}$. However, as $t$ approaches the value of the observation interval $\tau$, the decay deviates from the pure power law $1/t^{2+T}$, as the numerator in Eq.~(\ref{eq:psscaling}) can no longer be approximated by a constant. This deviation is solely a consequence of the finiteness of the observation interval. The scaling $P_{\textnormal{c}}(t) \propto 1/t^{2+T}$ is illustrated in Fig.~\ref{fig:val1}. Next, we analyze the intercontact distribution.
\vspace{0.5cm}

%%%%%%%%%%%%%%%%%%%%%%%%%%%%%%%%%%%%%%%%%%%%%%%%%%%%%%%%%%%%%%%%%%%%%%%%%%%%%%%%%%%%%%%%%%%%%%%%%%%%%%%%%%%%%%%%%%%%%%%
\section{Distribution of intercontact durations}
\label{sec:intercontact}

The intercontact distribution is dual to the contact distribution, and to derive it, we follow a similar procedure. Specifically, here we need to consider the probability of observing a sequence of exactly $t$ consecutive time slots where two nodes $i$ and $j$ with hidden degrees $\kappa_i$ and $\kappa_j$ and angular distance $\Delta \theta_{ij}$ are disconnected. Any such sequence should be enclosed within two slots where the two nodes are connected. Here we do not consider boundary cases, where the first or last of the $t$ slots starts or ends at the beginning or end of the observation period $\tau$, since by definition an intercontact duration should be enclosed within two contacts. Therefore, $t$ ranges from $1$ to $\tau-2$. We denote the above probability by $r_\textnormal{ic}(t; \kappa_i, \kappa_j, \Delta\theta_{ij})$.

We observe the following:
\begin{enumerate}
\item[(i)] The unconditional probability that two nodes $i$ and $j$ are connected in a slot $s$ is $p_{ij}$, where $p_{ij}$ is given by Eq.~(\ref{eq:p_s1}).
\item[(ii)] Given that they are connected in slot $s$, the probability that $i$ and $j$ are disconnected in slot $s+1$ is $(1-\omega_1)(1-\tilde{p}_{ij})$, where $\tilde{p}_{ij}$ is given by Eq.~(\ref{eq:p_s1omega}).
\item[(iii)] Given that they are disconnected in slot $s+1$, the probability that $i$ and $j$ remain disconnected in slots $s+2, \ldots, s+t$ is $[1 - (1-\omega_2)\tilde{p}_{ij} ]^{t-1}$.
\item[(iv)] Finally, given that they are disconnected in slot $s+t$, the probability that $i$ and $j$ are connected in slot $s+t+1$ is $(1-\omega_2)\tilde{p}_{ij}$.
\end{enumerate}
The probability $r_\textnormal{ic}(t;\kappa_i,\kappa_j,\Delta\theta_{ij})$ is obtained by multiplying $g_\tau(t)$ in Eq.~(\ref{eq:g_tau}) with the probabilities described in points (i) to (iv) above,
\begin{widetext}
\begin{equation}
\label{eq:p_ic}
r_\textnormal{ic}(t;\kappa_i,\kappa_j,\Delta\theta_{ij})=g_\tau(t)(1-\omega_1)(1-\omega_2)p_{ij}\tilde{p}_{ij}(1-\tilde{p}_{ij})[1 - (1-\omega_2)\tilde{p}_{ij} ]^{t-1}.
\end{equation}
\end{widetext}

The intercontact distribution, denoted as $P_{\textnormal{ic}}(t)$ and defined for $t\geq 1$, is given by
\begin{equation}
\label{eq:p_ic_norm}
P_{\textnormal{ic}}(t) = \frac{r_\textnormal{ic}(t)}{\sum_{j} r_\textnormal{ic}(j)} \propto r_\textnormal{ic}(t),
\end{equation}
where $r_\textnormal{ic}(t)$ is determined by removing the conditions on $\kappa_i$, $\kappa_j$, and $\Delta \theta_{ij}$ from Eq.~(\ref{eq:p_ic}),
\begin{widetext}
\begin{equation}
r_\textnormal{ic}(t) = \int \int \int r_\textnormal{ic}(t ; \kappa, \kappa', \Delta\theta) \rho(\kappa) \rho(\kappa') f(\Delta \theta) \mathrm{d} \kappa \mathrm{d} \kappa' \mathrm{d} \Delta\theta.
\end{equation}
\end{widetext}

\begin{figure*}
\includegraphics[width=3in]{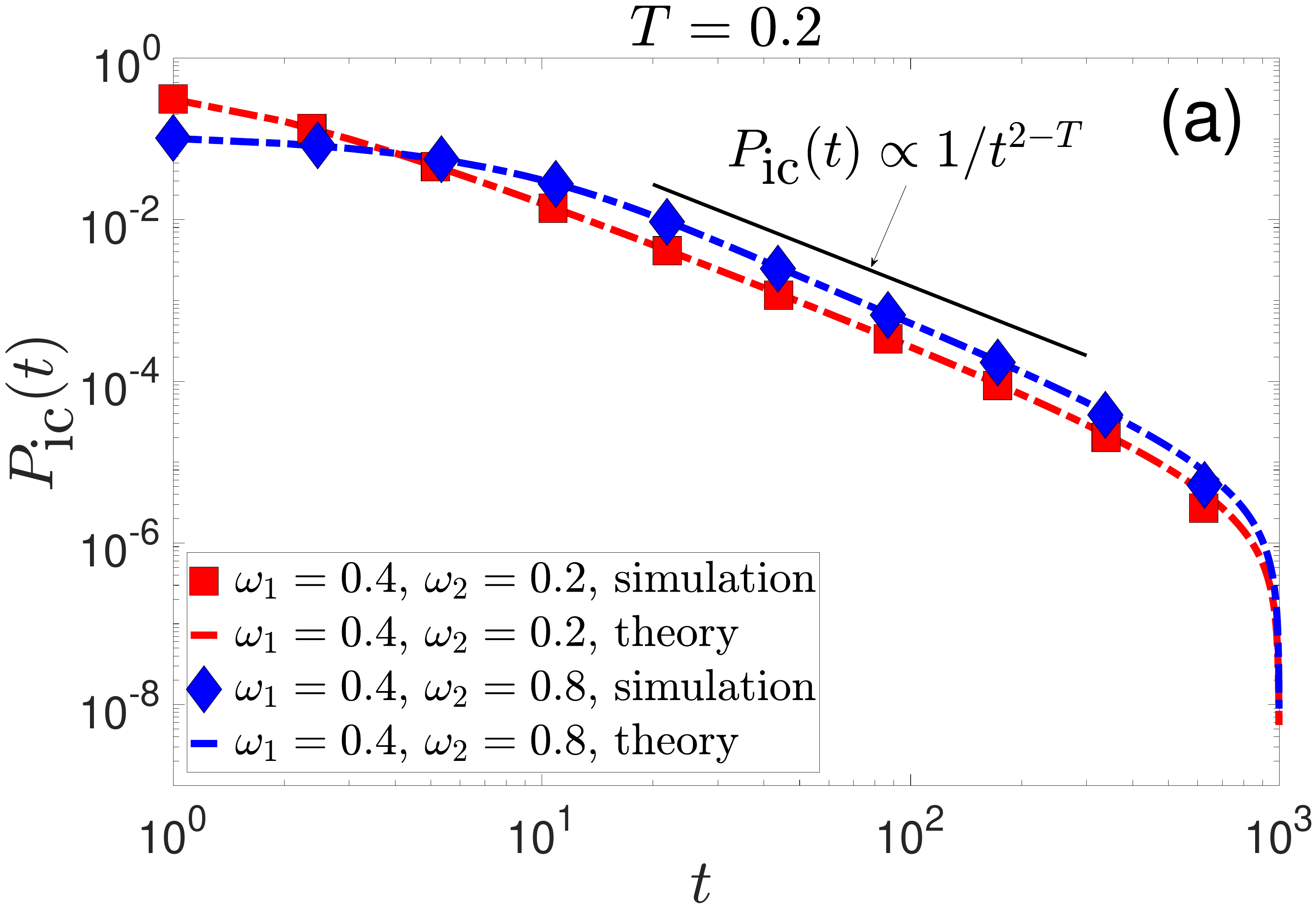}
\includegraphics[width=3in]{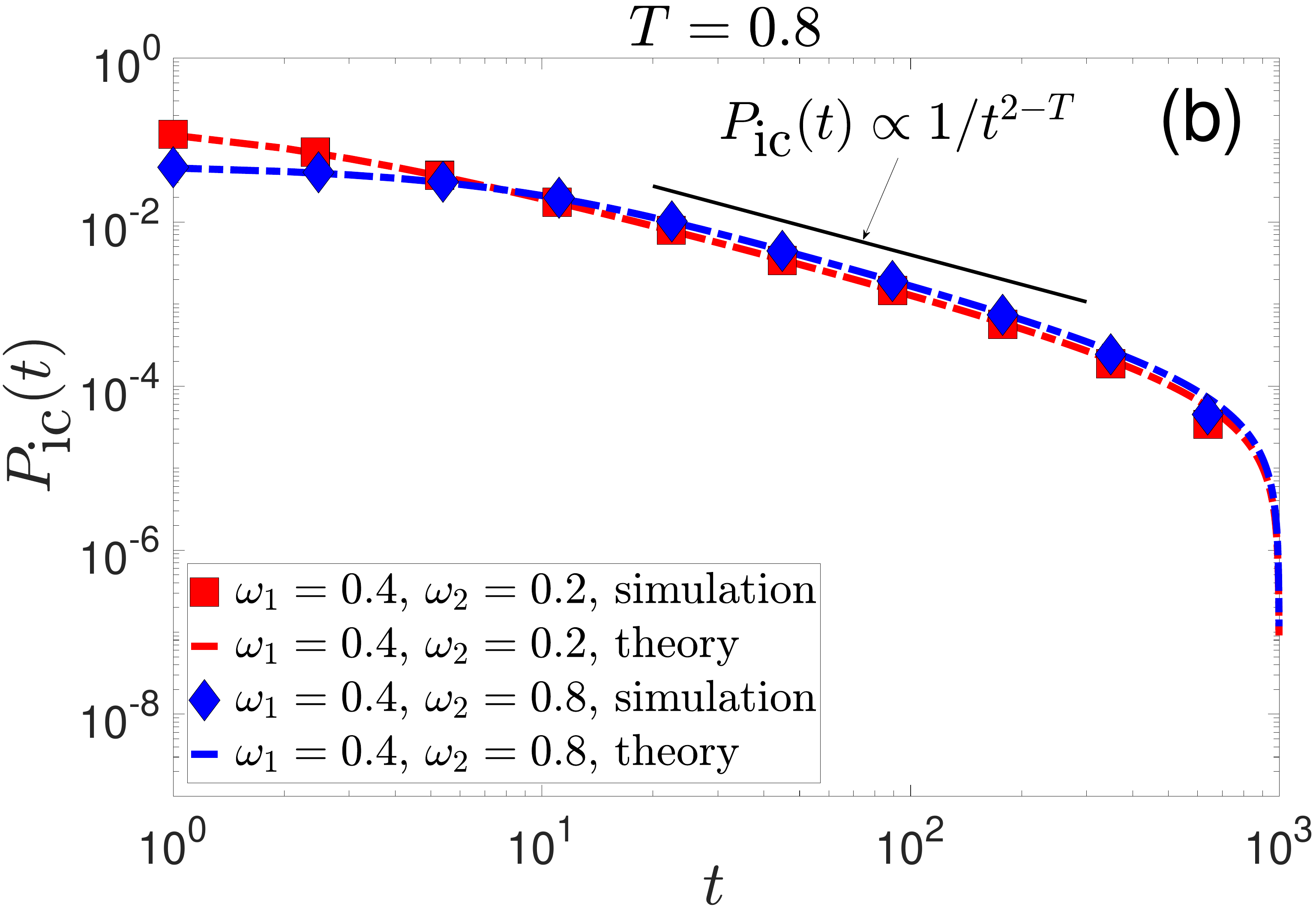}
\caption{Distribution of intercontact durations in simulated networks with the $(\omega_1, \omega_2)$-dynamic-$\mathbb{S}^{1}$ model~vs.~theoretical predictions. The latter are given by~$P_{\textnormal{ic}}(t)=r_\textnormal{ic}(t)/\sum_{j=1}^{\tau-2} r_\textnormal{ic}(j)$, where $r_\textnormal{ic}(t)$ is given by Eq.~(\ref{eq:ric_approx1}). Results are presented for two combinations of the persistence probabilities $\omega_1$ and $\omega_2$. All other simulation parameters are the same as in Fig.~\ref{fig:val1}.
\label{fig:val2}}
\end{figure*}

Following the same procedure as before to remove the conditions on  $\kappa_i$, $\kappa_j$, and $\Delta \theta_{ij}$, and employing the same transformations, we can write that for sufficiently large networks
\begin{widetext}
\begin{align}
\nonumber r_\textnormal{ic}(t) & \approx g_\tau(t) \frac{2\mu \bar{\kappa}^2 T}{N} (1-\omega_1)^{T}(1-\omega_2)^{2-T} \int_0^1 u^{1-T} (1-u)^{T}[1-(1-\omega_2)u]^{t-1}\Big(1-\frac{\omega_2-\omega_1}{1-\omega_1}u\Big)^{-1} \mathrm{d}u\\
&=g_\tau(t) \frac{\bar{k}T (1-T)}{2 N} (1-\omega_1)^{T}(1-\omega_2)^{2-T}F_1[2-T, 1-t, 1, 3; 1-\omega_2, \frac{\omega_2-\omega_1}{1-\omega_1}].
\label{eq:ric_approx1} 
\end{align}
\end{widetext}
We can observe the perfect duality between $r_\textnormal{ic}(t)$ and $r_\textnormal{c}(t)$, in the sense that Eq.~(\ref{eq:ric_approx1}) becomes Eq.~(\ref{eq:rc_appel}), if we exchange $\omega_2$ with $\omega_1$, $T$ with $-T$, and multiply the resulting relation by $-1$. The above analysis is validated in Fig.~\ref{fig:val2}.

\begin{figure*}
\includegraphics[width=2.89in]{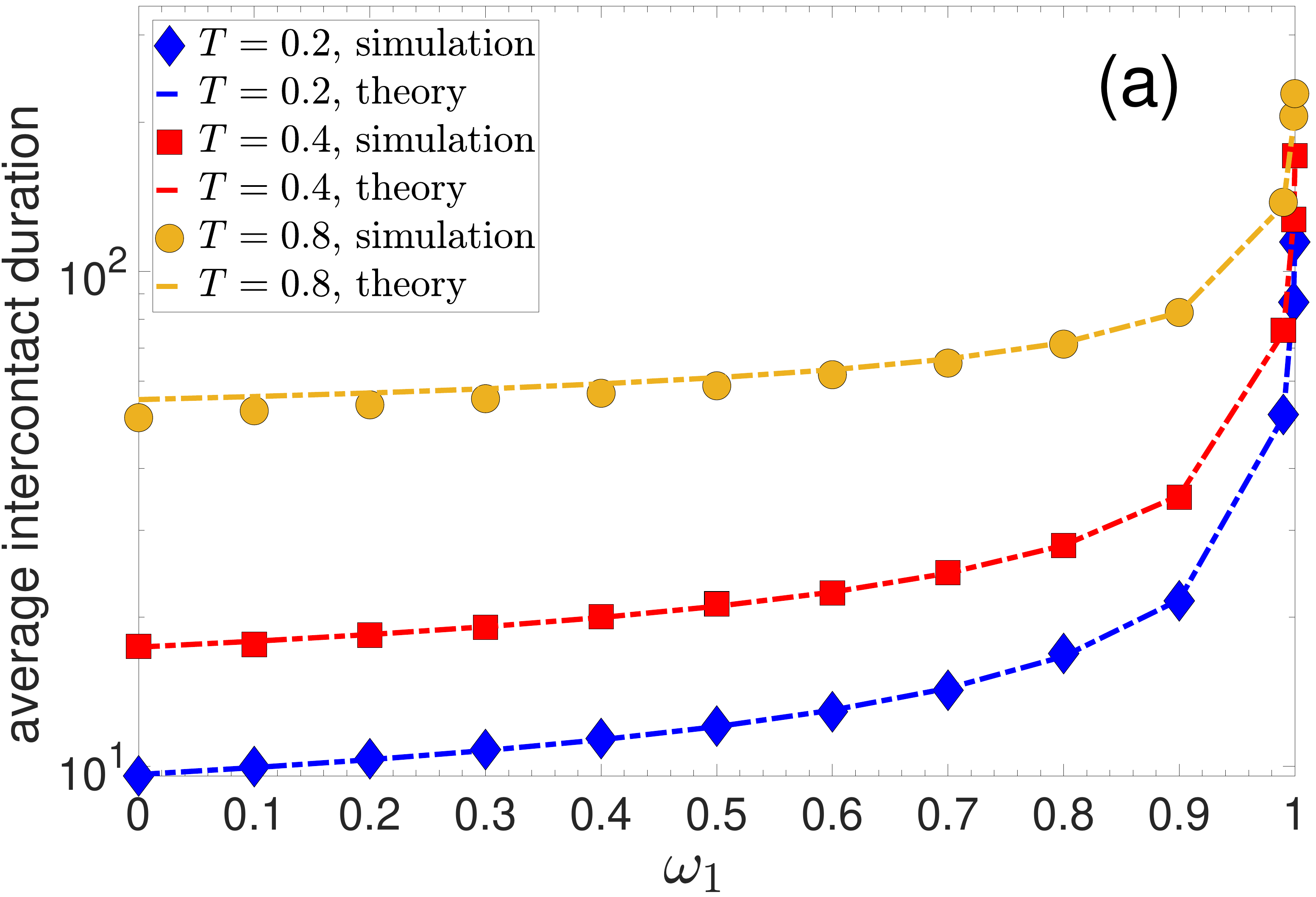}
\includegraphics[width=2.89in]{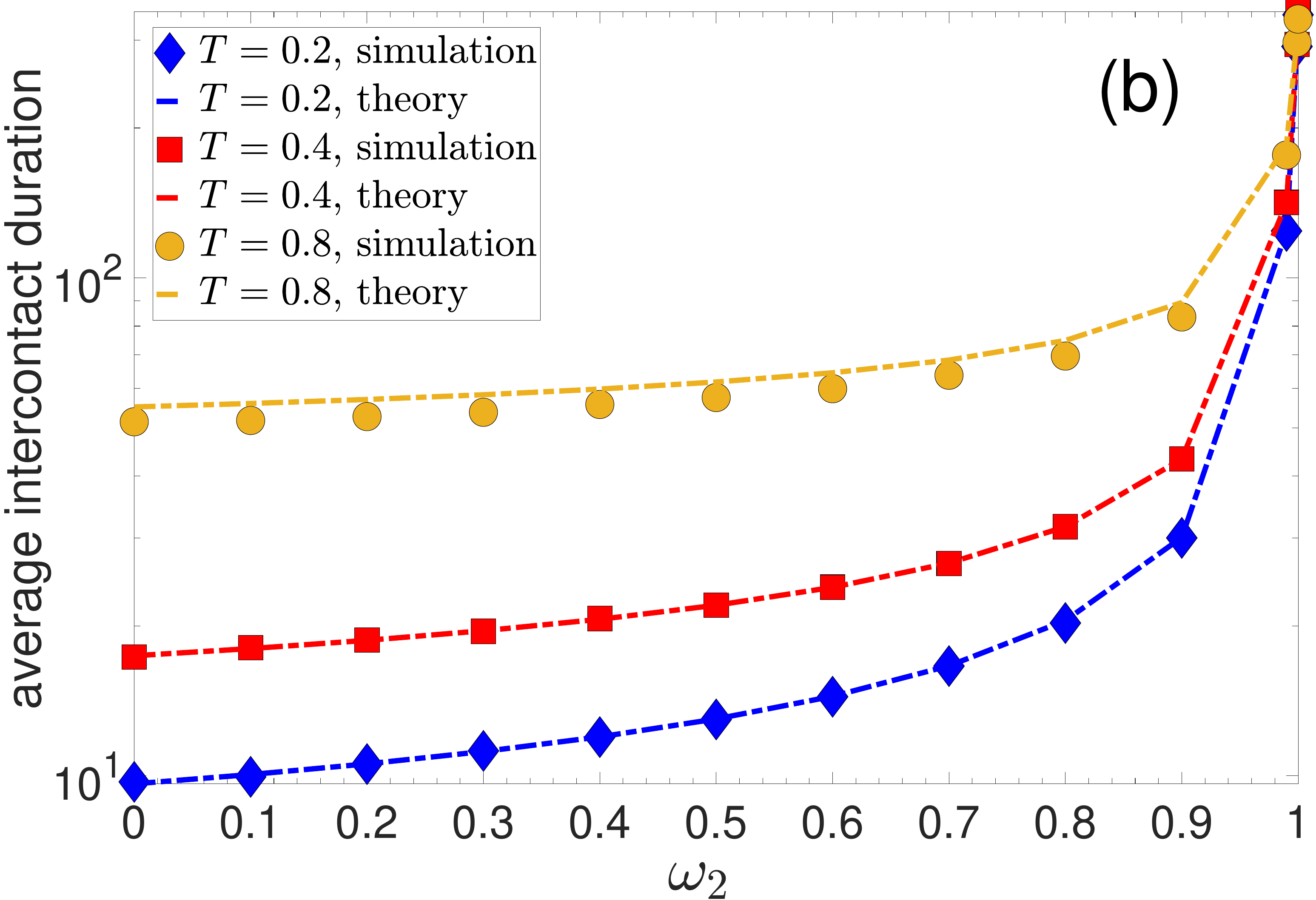}
\caption{Same as in Fig.~\ref{fig:ave_contact}, but for the average intercontact duration. The theoretical predictions (dashed lines) are given by $\bar{t}_\textnormal{ic}=\sum_{t=1}^{\tau-2} t P_\textnormal{ic}(t)$, where $P_\textnormal{ic}(t)$ is computed as in Fig.~\ref{fig:val2}.
\label{fig:ave_intercontact}}
\end{figure*}

%%%%%%%%%%%%%%%%%%%%%%%%%%%%%%%%%%%%%%%%%%%%%%%%%%%%%%%%%%%%%%%%%%%%%%%%%%%%%%%%%%%%%%%%%%%%%%%%%%%%%%%%%%%%%%%%%%%%%%%%
\emph{Average intercontact duration.} In Fig.~\ref{fig:ave_intercontact}, we investigate how  parameters $\omega_1$, $\omega_2$, and $T$ affect the average intercontact duration. As with the case of the average contact duration, we see that the average intercontact duration also increases with $\omega_1$ or $\omega_2$, with the rate of increase becoming more pronounced as these parameters approach $1$. Further, the increase occurs faster with $\omega_2$ than with $\omega_1$, especially as these parameters approach $1$. This is expected, as $\omega_2$ directly impacts the probability that two nodes remain disconnected via Eq.~(\ref{eq:cp2b}). It can be shown that as $\omega_2$ approaches $1$, $P_\textnormal{ic}(t)$ becomes proportional to $g_\tau(t)$, and the average intercontact duration tends to $\tau/3$. On the other hand, as $\omega_1$ approaches $1$, $P_\textnormal{ic}(t)$ becomes proportional to $g_\tau(t) {_2}F_1[1-T, 1-t, 2; 1-\omega_2]$, while the average intercontact duration is upper-bounded by $\tau/3$. The average intercontact duration also increases with $T$, while remaining upper-bounded by $\tau/3$. This is because higher values of $T$ increase randomness in the connections, thereby reducing the probability of pairs reconnecting. We note that intercontacts cannot be defined for $\omega_1$ or $\omega_2$ exactly equal to 1, or for $T = 0$, as in these cases there are no link dynamics.

%%%%%%%%%%%%%%%%%%%%%%%%%%%%%%%%%%%%%%%%%%%%%%%%%%%%%%%%%%%%%%%%%%%%%%%%%%%%%%%%%%%%%%%%%%%%%%%%%%%%%%%%%%%%%%%%%%%%%%%%
\emph{Tail of the intercontact distribution.} Finally, given the duality between Eqs.~(\ref{eq:ric_approx1}) and~(\ref{eq:rc_appel}), we can follow exactly the same procedure as in the case of Eq.~(\ref{eq:rc_appel}), to show that for large $t$, $r_\textnormal{ic}(t)$ can be approximated as
\begin{equation}
r_\textnormal{ic}(t) \approx g_\tau(t) \frac{\bar{k} T (1-T) (1-\omega_1)^{T}}{N \Gamma(1+T)}\frac{1}{t^{2-T}}\propto \frac{g_\tau(t)}{t^{2-T}}.
\end{equation}
The above result holds true for any combination of $\omega_1, \omega_2 \in [0,1)$. For $t \ll \tau$, $g_\tau(t)\approx 1$, and thus $r_\textnormal{ic}(t)$, and consequently, the intercontact distribution $P_{\textnormal{ic}}(t)$, decay according to the power law $1/t^{2-T}$. The scaling $P_{\textnormal{ic}}(t) \propto 1/t^{2-T}$ is illustrated in Fig.~\ref{fig:val2}. In the next section, we turn our attention to the expected time-aggregated degree.

%%%%%%%%%%%%%%%%%%%%%%%%%%%%%%%%%%%%%%%%%%%%%%%%%%%%%%%%%%%%%%%%%%%%%%%%%%%%%%%%%%%%%%%%%%%%%%%%%%%%%%%%%%%%%%%%%%%%%%%%
\section{Time-aggregated degree}
\label{sec:degree}

To analyze the expected time-aggregated degree, we need to consider the probability that two nodes $i$ and $j$ with hidden degrees $\kappa_i$ and $\kappa_j$ and angular distance $\Delta\theta_{ij}$ do not connect during the observation period $\tau$. This probability is given by
\begin{equation}
\label{eq:r_0_cond}
r_0(\kappa_i, \kappa_j, \Delta\theta_{ij}) = (1-p_{ij}) [1 - (1-\omega_2) \tilde{p}_{ij}]^{\tau-1},
\end{equation}
where $p_{ij}$ and $\tilde{p}_{ij}$ are given by Eqs.~(\ref{eq:p_s1}) and~(\ref{eq:p_s1omega}).

The expected time-aggregated degree, denoted as $\bar{k}_\textnormal{aggr}$, is given by
\begin{equation}
\label{eq:kaggr}
\bar{k}_\textnormal{aggr}=(N-1)(1-r_0),
\end{equation}
where $r_0$ is determined by removing the conditions on $\kappa_i$, $\kappa_j$, and $\Delta \theta_{ij}$ from Eq.~(\ref{eq:r_0_cond}),
\begin{equation}
\label{eq:r_0}
r_0 = \int \int \int r_0(\kappa, \kappa', \Delta\theta) \rho(\kappa) \rho(\kappa') f(\Delta \theta) \mathrm{d} \kappa \mathrm{d} \kappa' \mathrm{d} \Delta\theta.
\end{equation}

Following the same procedure as before to remove the condition on $\Delta \theta_{ij}$, we can write
\begin{widetext}
\begin{equation}
\label{eq:r0kappa} 
r_0(\kappa_i, \kappa_j)=\frac {2\mu\kappa_i\kappa_j T}{N} \Big(\frac{1-\omega_1}{1-\omega_2}\Big)^T\int_{u_0^{ij}}^1 u^{-(1+T)}(1-u)^T [1-(1-\omega_2) u]^{\tau-1}\Big(1-\frac{\omega_2-\omega_1}{1-\omega_1}u\Big)^{-1}\mathrm{d}u,
\end{equation}
\end{widetext}
where $u_0^{ij}$ is as in Eq.~(\ref{eq:full_contact_integral}). 

\begin{figure*}
\includegraphics[width=2.9in]{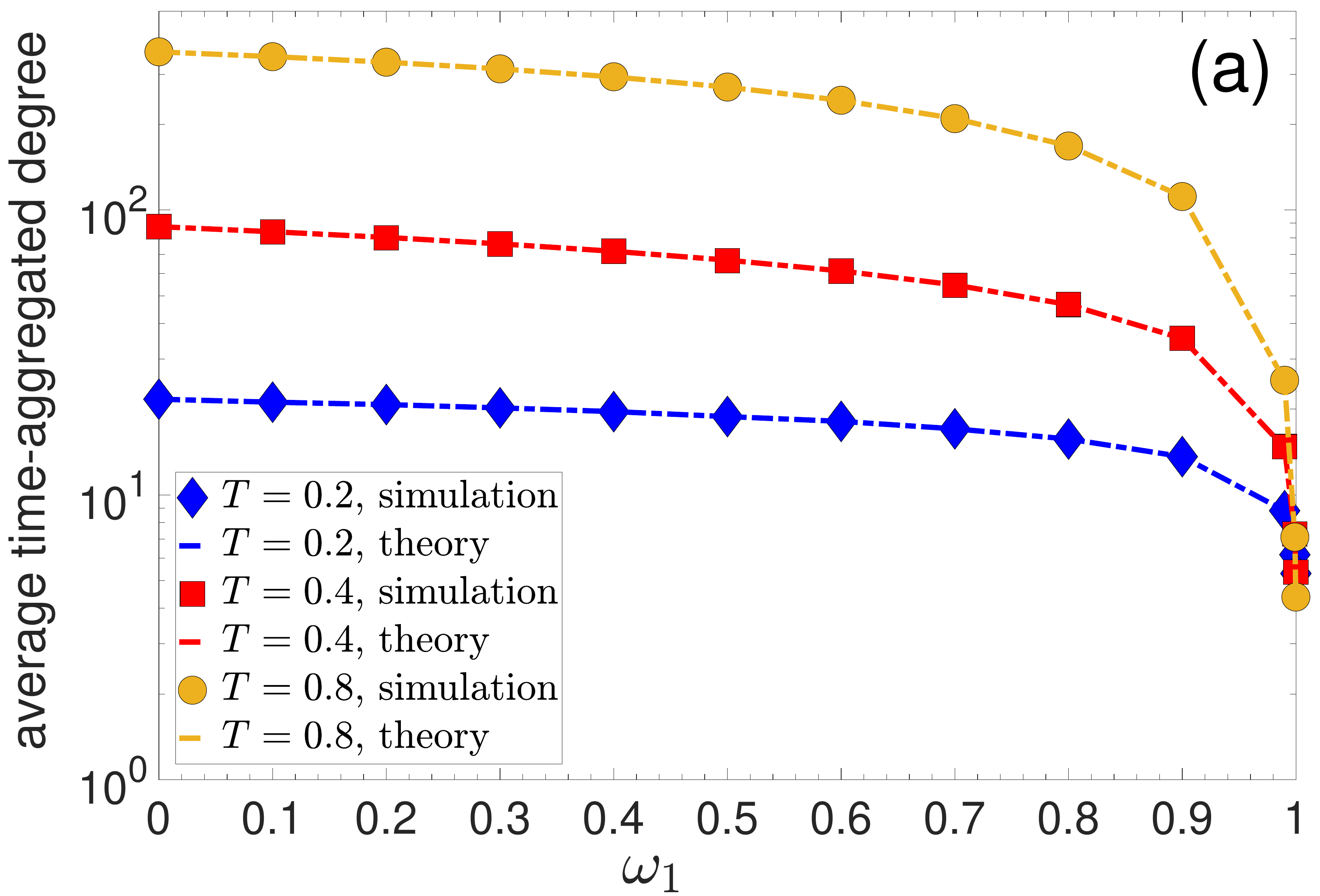}
\includegraphics[width=2.9in]{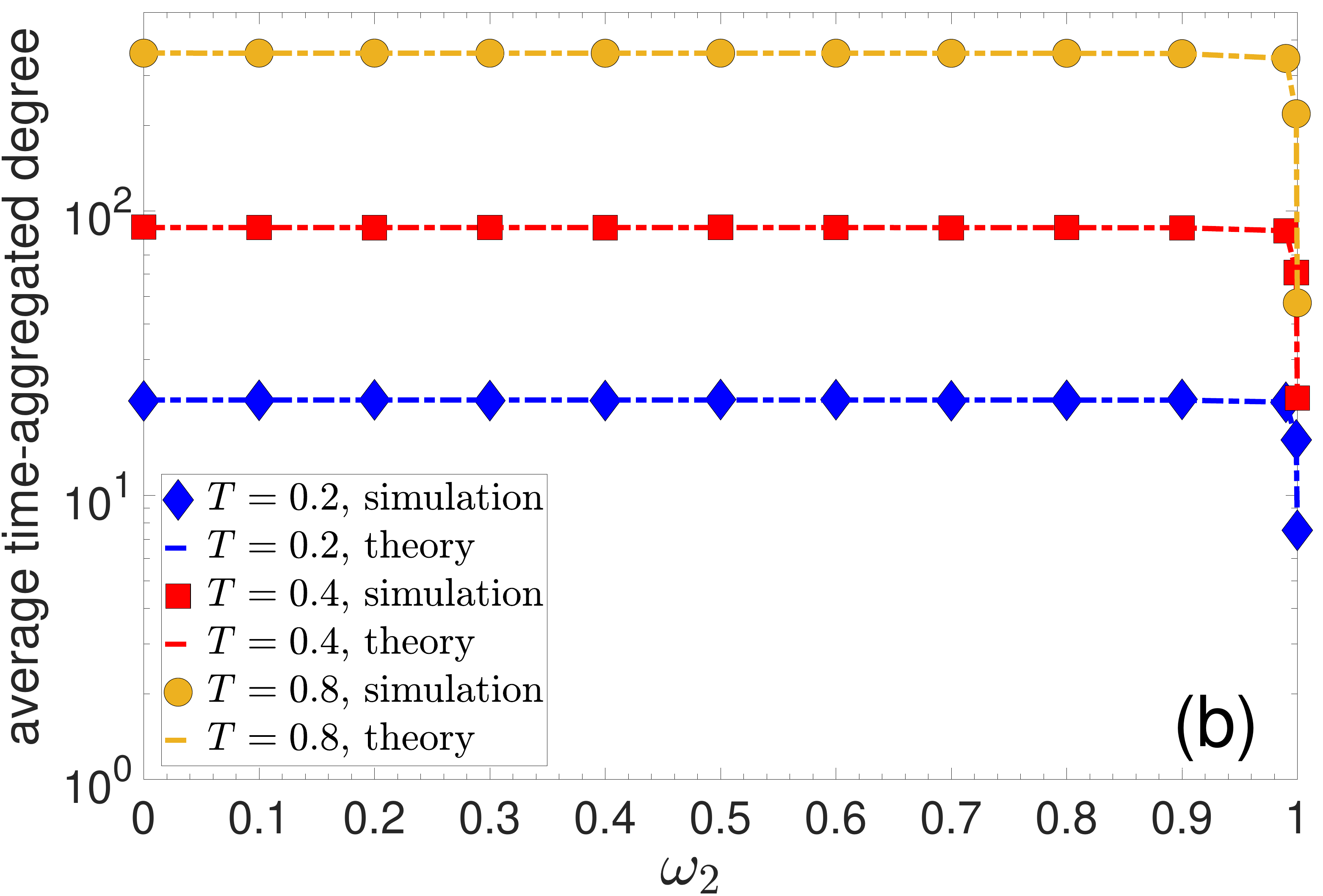}
\caption{Average time-aggregated degree vs. $\omega_1$, $\omega_2$, and $T$. Plot (a) shows the average time-aggregated degree as a function of the persistence probability of connections $\omega_1$. The persistence probability of disconnections, $\omega_2$, is set to zero. Results are shown for different values of the network temperature $T$. In each case the three rightmost points correspond respectively to $\omega_1 = 0.99$, $0.999$, and $0.9999$. All other parameters are the same as in Fig.~\ref{fig:val1}. The dashed lines depict theoretical predictions given by Eqs.~(\ref{eq:kaggr})-(\ref{eq:r0kappa}). Plot (b) is similar to (a), except that $\omega_1$ is set to zero, and we vary $\omega_2$. The $y$-axes use a logarithmic scale.
\label{fig:kaggr}}
\end{figure*}

The integral in Eq.~(\ref{eq:r0kappa}) diverges for $N \to \infty$, i.e., for $u_0^{ij}\to 0$. Therefore, we cannot consider its ``large-$N$ approximation'' by setting $u_0^{ij}=0$ as its lower limit. In particular, as shown for the case of $\omega_1=\omega_2$, $\bar{k}_\textnormal{aggr}$ is sensitive to finite size effects, especially at larger network temperatures~\cite{Papadopoulos2019, Zambirinis2022}, and to accurately compute it in general one needs to numerically evaluate the integrals in Eqs.~(\ref{eq:r0kappa}) and~(\ref{eq:r_0}). 

The above analysis is validated in Fig.~\ref{fig:kaggr}. We see from the figure that $\bar{k}_\textnormal{aggr}$ decreases as the link persistence probability $\omega_1$ increases, or as the network temperature $T$ decreases. In particular, as $\omega_1$ approaches $1$ or $T$ approaches $0$, $\bar{k}_\textnormal{aggr}$ converges to the average snapshot degree $\bar{k}$. Further, we see that $\bar{k}_\textnormal{aggr}$ remains virtually unaffected by the non-link persistence probability $\omega_2$, unless $\omega_2$ is very close to $1$. In particular, at the limit $\omega_2 \to 1$, $\bar{k}_\textnormal{aggr}$ tends again to $\bar{k}$. This explains why the performance of epidemic spreading processes may not be significantly affected by non-link persistence, unless it is very strong, cf.~Appendix~\ref{sec:appendix0}.

Table~\ref{tab:depsum} provides a summary of how $\bar{k}_\textnormal{aggr}$, as well as the average contact and intercontact durations ($\bar{t}_\text{c}$ and $\bar{t}_\text{ic}$) change with parameters $T$, $\omega_1$, and $\omega_2$.
\begin{table}
\centering
\begin{tabularx}{\columnwidth}{*{4}{>{\centering\arraybackslash}X}}
\multicolumn{1}{c}{} & $\bar{k}_{\text{aggr}}$ & $\bar{t}_\text{c}$ & $\bar{t}_\text{ic}$ \\
\hline
$T$ & $\nearrow_\star$ & $\searrow$ & $\nearrow_\star$ \\
\hline
$\omega_1$ & $\searrow_\star$ & $\nearrow_\star$ & $\nearrow$ \\
\hline
$\omega_2$ & $\searrow$ & $\nearrow$ & $\nearrow_\star$ \\
\hline
\end{tabularx}
\caption{Summary of dependencies of $\bar{k}_{\text{aggr}}$, $\bar{t}_\text{c}$, and $\bar{t}_\text{ic}$, on parameters $T$, $\omega_1$, and $\omega_2$. Arrows indicate an increase ($\nearrow$) or decrease ($\searrow$) of the corresponding average as $T$, $\omega_1$, or $\omega_2$ increases. Stars indicate the averages that generally change more rapidly with a change in the corresponding parameter.}
\label{tab:depsum}
\end{table}

%%%%%%%%%%%%%%%%%%%%%%%%%%%%%%%%%%%%%%%%%%%%%%%%%%%%%%%%%%%%%%%%%%%%%%%%%%%%%%%%%%%%%%%%%%%%%%%%%%%%%%%%%%%%%%%%%%%%%%%%%%%%
\section{Other related work and discussion}
\label{sec:other_work}

In this section, we discuss our model in the context of other related work.

A popular model for temporal networks is the activity-driven model (ADM), introduced in Ref.~\cite{Perra2012} and extended to include node attractiveness in Ref.~\cite{Alessandretti2017}. The ADM has been regularly utilized due to its simplicity and adaptability, cf.~\cite{Sun2015, Perra2017, Nadini2018, Li2020, Cai2024}. However, it is not a geometric network model. In contrast, we generalize temporal network modeling based on RHGs, which have been shown to naturally reflect real-world networks~\cite{Krioukov2010, Papadopoulos2012, Boguna2010, Papadopoulos2019, Zambirinis2022}. Additionally, while ADM analyses have primarily focused on properties of the time-aggregated network, such as its degree distribution~\cite{Perra2012, StarniniADM2013}, our work focuses on properties of the resulting temporal network itself, such as its (inter)contact distributions.

Other methodologies have extended popular static network models, such as Erdős–Rényi (ER) random graphs, the configuration model, the stochastic block model, and models with hidden variables, to temporal settings~\cite{Moore2017, mazzarisi2020, friel2016}. These approaches account for link and non-link persistence with different rates in a Markovian manner, similar to our work. However, they do not consider geometric network models or models where the node hidden variables represent their popularity and similarity coordinates in an underlying hyperbolic space.

Non-Markovian link persistence has also been considered, cf.~\cite{Williams2019}. Additionally, the work in Ref.~\cite{hartle2021} investigated the interplay between hidden variable dynamics and link dynamics in temporal network models. The $\omega$-dynamic-$\mathbb{S}^1$ model~\cite{Zambirinis2022} is a special case of the general class of models discussed in Ref.~\cite{hartle2021}, where there are no hidden variable dynamics.

Moreover, a substantial body of work has studied the effects of temporality on various dynamical processes, including epidemic spreading~\cite{vazquez2007, timo2009, Karsai2011, machens2013, gauvin2013, Granell2018, Williams2019}, synchronization and diffusion~\cite{Masuda2013}, the evolution of cooperation~\cite{Li2020}, and the emergence of chaos~\cite{Rock2023}. Often, simple null models are utilized in such studies, such as the ADM~\cite{Li2020} or models based on random graphs~\cite{Williams2019}. The ($\omega_1, \omega_2$)-dynamic-$\mathbb{S}^1$ constitutes an important addition to the suite of such models. The model is based on a principled geometric framework (RHGs), yields realistic dynamical properties, and allows simultaneous control of (i) the expected degree distribution in the snapshots via $\rho(\kappa)$, (ii) the localization of connections and thereby clustering via $T$, and (iii) the stability of connections and disconnections via $\omega_1$ and $\omega_2$.

Fully investigating the effects and interplay of the model's parameters on different dynamical processes is beyond the scope of this paper. However, we have considered some illustrative examples (for certain settings of the model's parameters) in the context of epidemic spreading (Figs.~\ref{fig:epidemics} and~\ref{fig:epidemics_app}). These examples demonstrate that link and non-link persistence can slow down spreading, depending on the setting and the network temperature $T$. The work in Ref.~\cite{Williams2019} also observed that link persistence can slow down spreading, utilizing a model based on ER random graphs. However, ER random graphs correspond to the limit $T \to \infty$ in RHGs, where the nodes' popularity and similarity coordinates are completely ignored~\cite{Krioukov2010}. Finally, the observation that increasing clustering (by decreasing $T$) can also suppress overall spreading is intuitive and in line with prior work~\cite{Szendri2004,Salathe2010}.

%%%%%%%%%%%%%%%%%%%%%%%%%%%%%%%%%%%%%%%%%%%%%%%%%%%%%%%%%%%%%%%%%%%%%%%%%%%%%%%%%%%%%%%%%%%%%%%%%%%%%%%%%%%%%%%%%%%%%%%%

\section{Conclusion}
\label{sec:conclusion}
We have generalized temporal random hyperbolic graphs by introducing distinct probabilities $\omega_1$ and $\omega_2$ for link and non-link persistence, and elucidated the non-trivial dependence of key temporal network properties on link and non-link persistence strength, and on the network temperature $T$. The generalized model can be used to study a wider range of scenarios involving dynamical processes on temporal networks. This is because it allows more flexible tuning of the average contact and intercontact durations, and of the average time-aggregated degree. Specifically, these quantities are now controlled by three parameters ($\omega_1, \omega_2$, $T$) instead of two ($\omega$, $T$). 

We have also proven that the tails of the contact and intercontact distributions decay as power laws with exponents $2+T$ and $2-T$, respectively, as in the case of $\omega_1=\omega_2$~\cite{Zambirinis2022}. An outstanding question is whether there exists a simple model extension in which the tails of these distributions are not coupled by the common parameter $T$, but can be tuned more independently. Another question is whether there exist model extensions in which the (inter)contact distributions deviate from pure power laws, as may be observed in real-world systems. Further, it may be worth investigating whether incorporating link persistence affects the conclusions about the non-realism of temporal RHGs in the hot regime ($T >1$), which has been analyzed in the absence of link persistence~\cite{Papadopoulos2022}.

Other interesting directions for future work include the inference of link and non-link persistence probabilities in real networks~\cite{Papadopoulos2019, papaefthymiou2024}, the derivation and analysis of models of temporal RHGs in higher dimensions~\cite{budel2023}, temporal RHG models with non-Markovian link persistence~\cite{Williams2019}, models where different pairs of nodes can have different link and non-link persistence probabilities~\cite{mazzarisi2020}, as well as temporal RHG models for bipartite networks~\cite{KPK2017, friel2016}.

%%%%%%%%%%%%%%%%%%%%%%%%%%%%%%%%%%%%%%%%%%%%%%%%%%%%%%%%%%%%%%%%%%%%%%%%%%%%%%%%%%%%%%%%%%%%%%%%%%%%%%%%%%%%%%%%%%%%%%%%%%%
\begin{acknowledgements}
The authors acknowledge support by the TV-HGGs project (OPPORTUNITY/0916/ERC-CoG/0003), co-funded by the European Regional Development Fund and the Republic of Cyprus through the Research and Innovation Foundation.
\end{acknowledgements}
%%%%%%%%%%%%%%%%%%%%%%%%%%%%%%%%%%%%%%%%%%%%%%%%%%%%%%%%%%%%%%%%%%%%%%%%%%%%%%%%%%%%%%%%%%%%%%%%%%%%%%%%%%%%%%%%%%%%%%%%%%%%
\appendix

\section{Epidemic spreading simulations} 
\label{sec:appendix0}
In Fig.~\ref{fig:epidemics}, we consider the Susceptible-Infected-Susceptible (SIS) and the Susceptible-Infected-Recovered (SIR) epidemic spreading models~\cite{sis_ref}. In the SIS model, each node can be in one of two states: susceptible (S) or infected (I). In each time slot, an infected node can recover with probability $\beta$ and become susceptible again, whereas infected nodes can infect the susceptible nodes they are connected to with probability $\alpha$. Thus, the transition of states is S$\rightarrow$I$\rightarrow$S. In the SIR model, each node can be in one of three states: susceptible (S), infected (I), or recovered (R). In each time slot, an infected node can recover with probability $\beta$, whereas infected nodes can infect the susceptible nodes they are connected to with probability $\alpha$. Thus, the transition of states is S$\rightarrow$I$\rightarrow$R. We note that nodes that get infected in a time slot will not attempt to infect susceptible neighbors until the next time slot. Also, in the case of SIS, nodes that recover in a time slot are not considered for infection until the next time slot. 

As mentioned in the caption of Fig.~\ref{fig:epidemics}, all simulations start with 5\% of the nodes randomly infected, i.e., in the I~state, while $\alpha=0.5$ and $\beta=0.005$. In each time slot, the network snapshots change according to the $(\omega_1, \omega_2)$-dynamic-$\mathbb{S}^{1}$ model, i.e., according to Eqs.~(\ref{eq:cp1b}) and~(\ref{eq:cp2b}). Therefore, the simulated SIS and SIR processes evolve at the same time scale as the simulated networks. 

\begin{figure*}[t]
\includegraphics[width=2.3in]{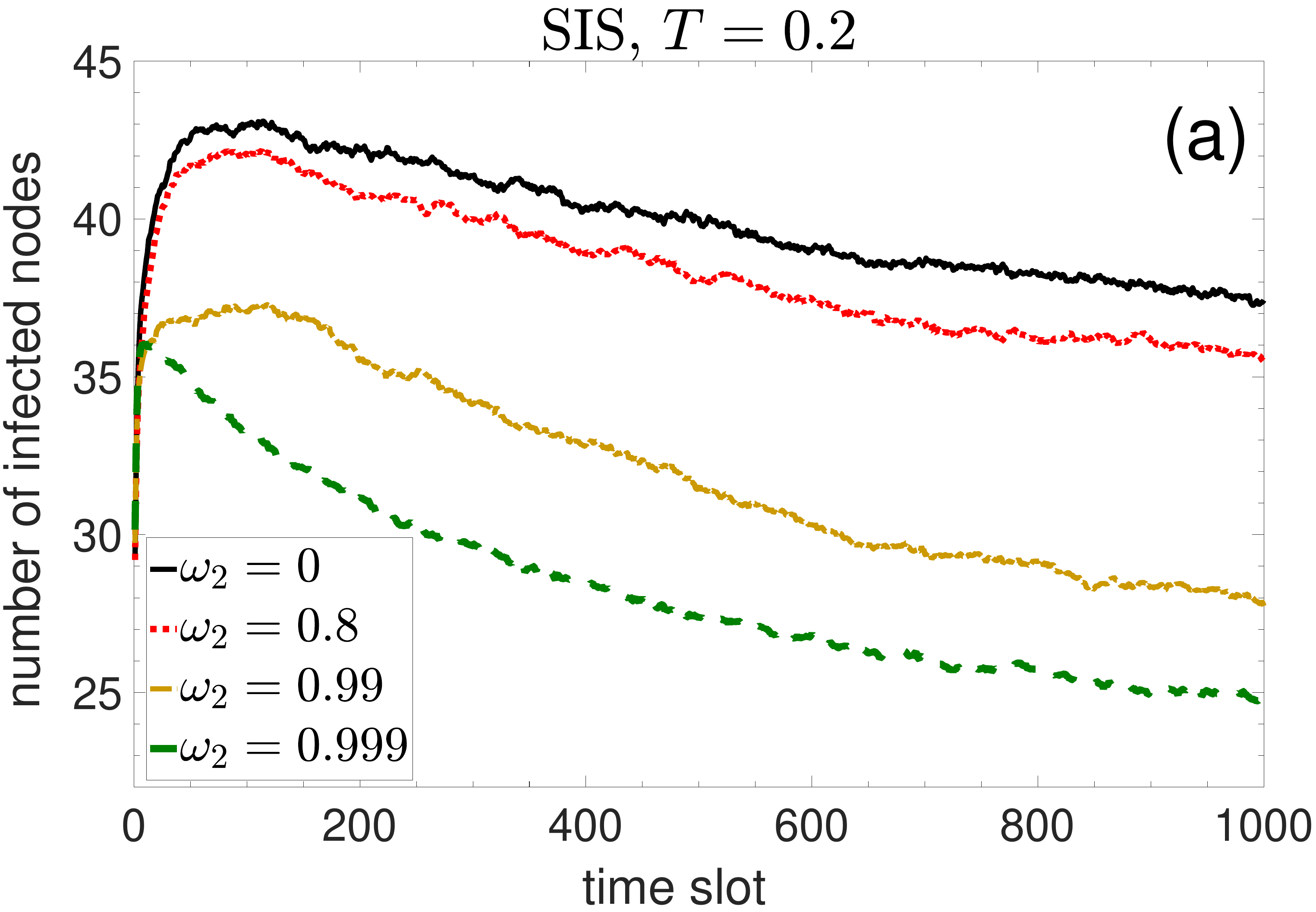}
\includegraphics[width=2.3in]{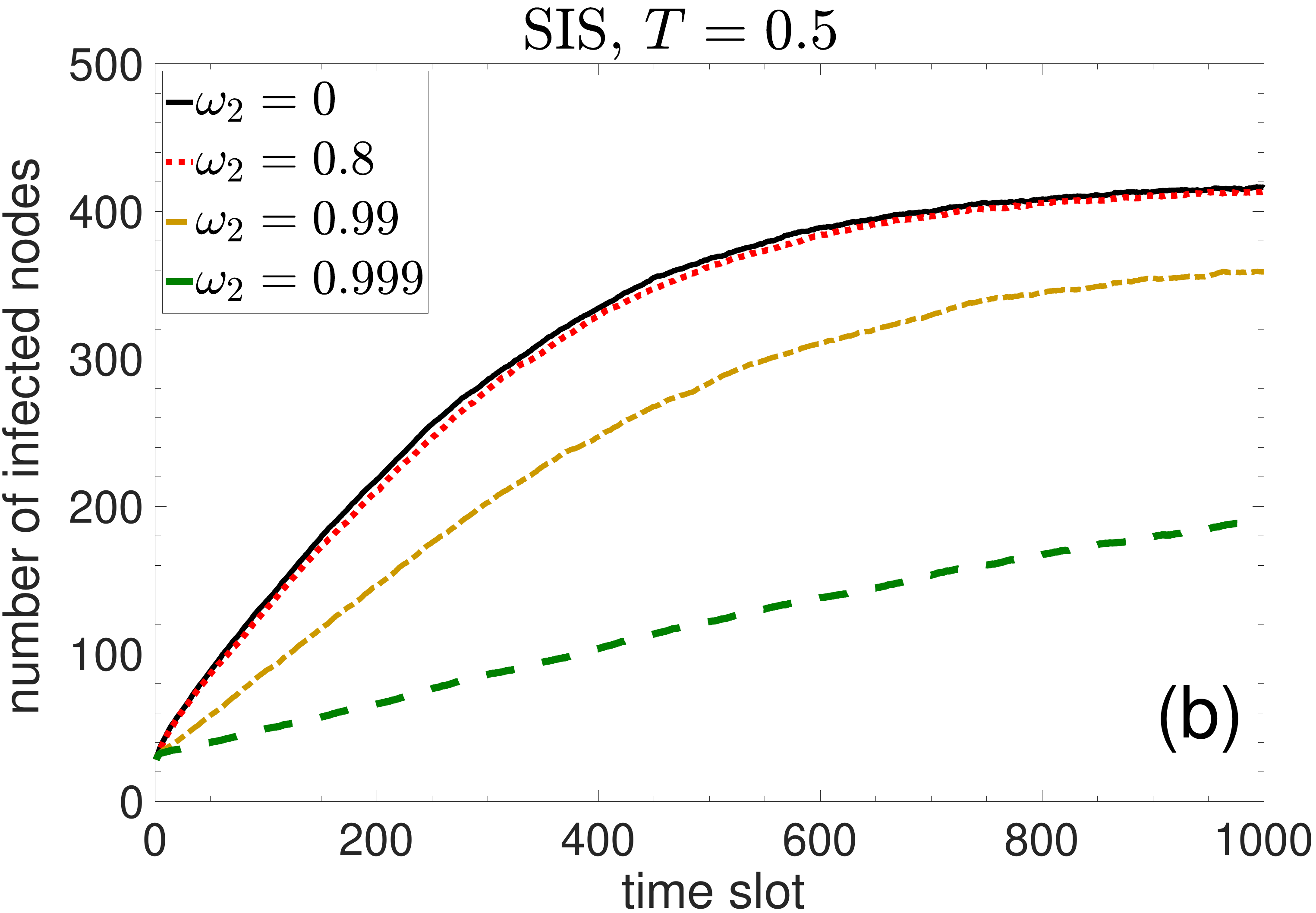}
\includegraphics[width=2.3in]{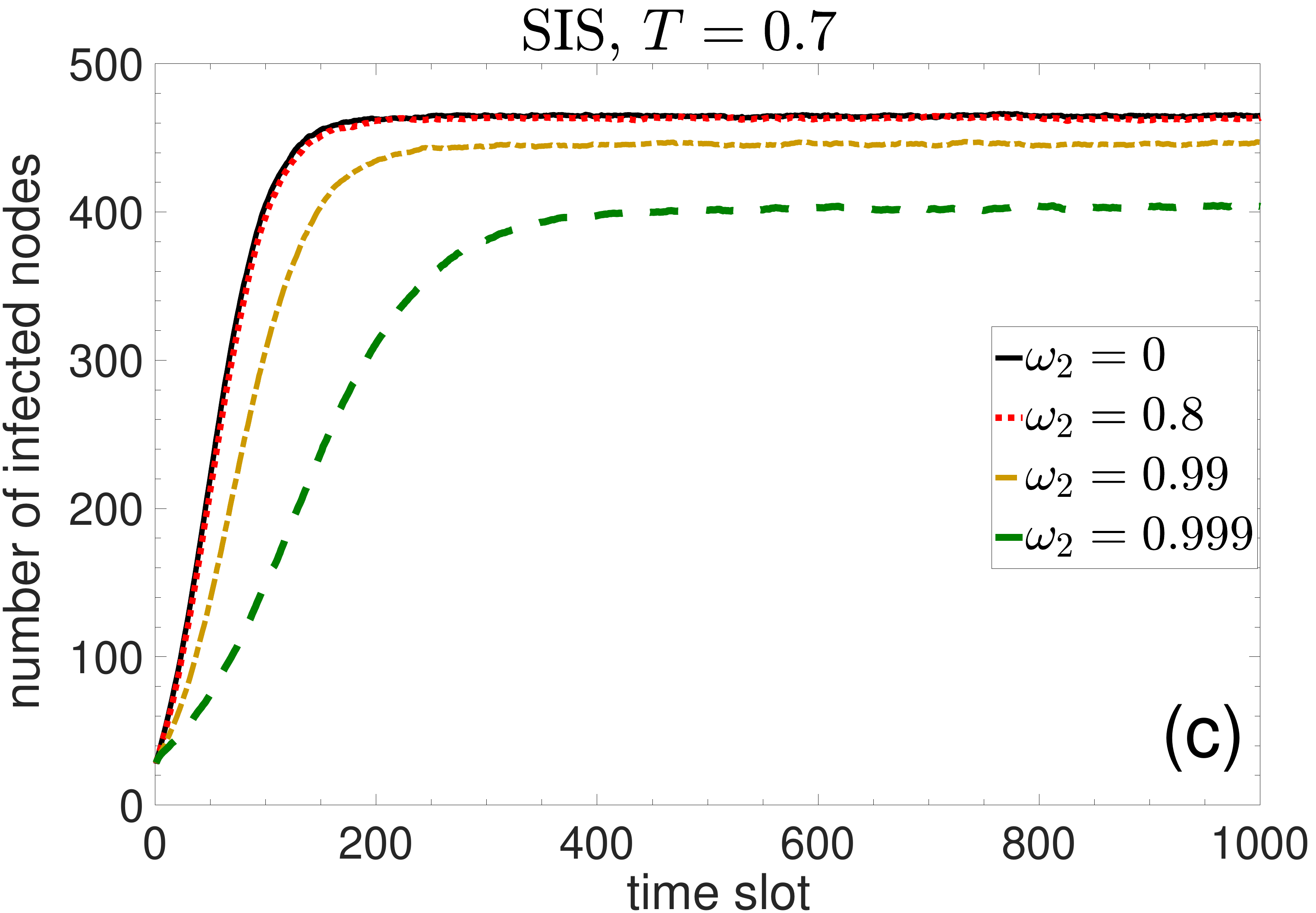}
\newline
\newline
\includegraphics[width=2.3in]{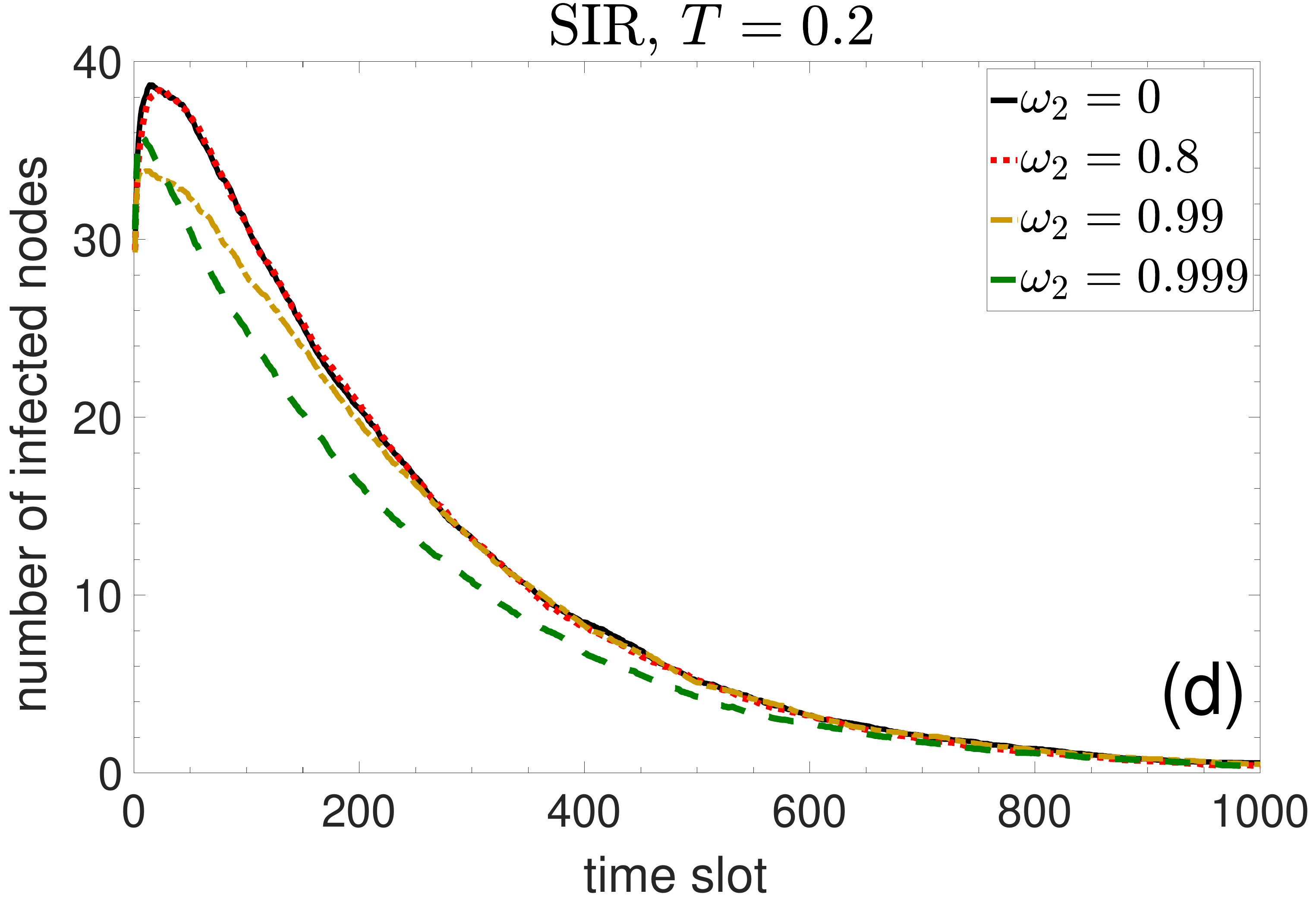}
\includegraphics[width=2.3in]{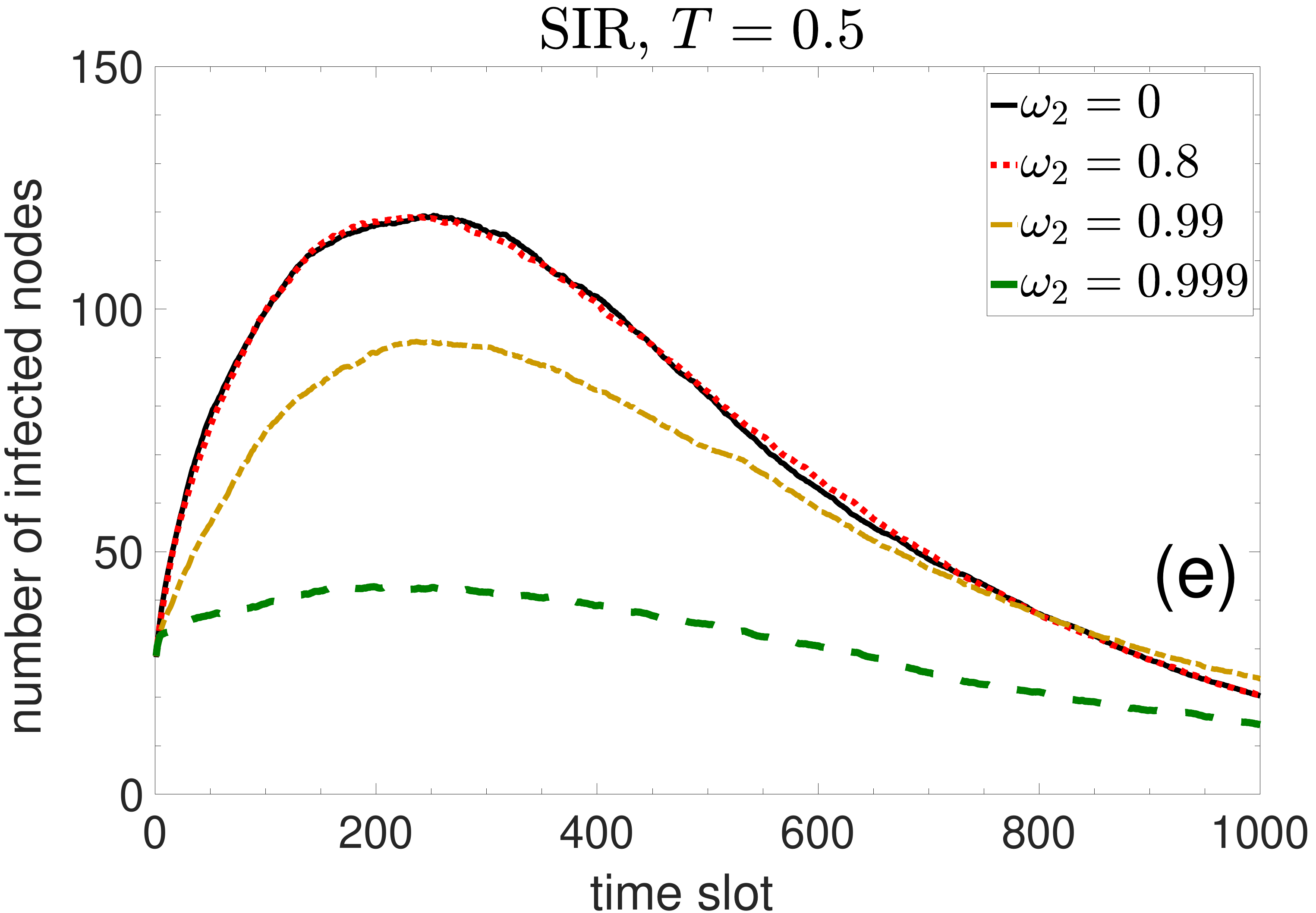}
\includegraphics[width=2.3in]{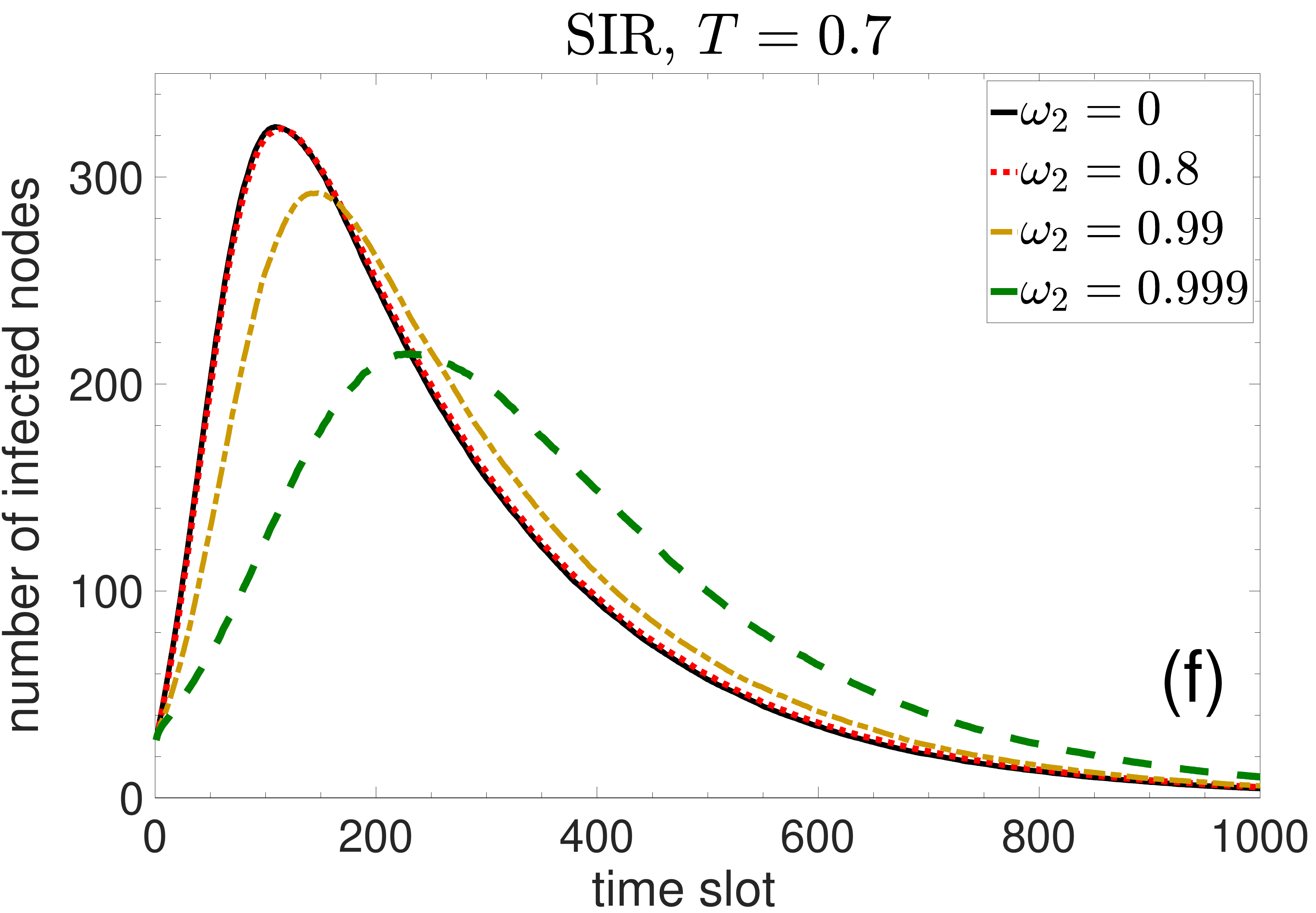}
\caption{Same as in Fig.~\ref{fig:epidemics}, except that results are presented for different levels of the non-link persistence probability $\omega_2$, while in all cases $\omega_1=0$. 
\label{fig:epidemics_app}}
\end{figure*}

\emph{Effect of non-link persistence.} Figure~\ref{fig:epidemics} illustrates the effect of link persistence in isolation from non-link persistence by setting $\omega_2 = 0$ and varying $\omega_1$. Conversely, Fig.~\ref{fig:epidemics_app} shows the effect of non-link persistence in isolation from link persistence by setting $\omega_1 = 0$ and varying $\omega_2$. As seen in Sec.~\ref{sec:degree}, non-link persistence has a much lesser effect on the expected time-aggregated degree compared to link persistence. However, its effect can become significant as $\omega_2$ approaches 1 [Fig.~\ref{fig:kaggr}(b)]. This is reflected in the performance of epidemic spreading in Fig.~\ref{fig:epidemics_app}, where $\omega_2$ needs to be very close to 1 to observe similarly notable differences as those seen in Fig.~\ref{fig:epidemics} with lower values of $\omega_1$.

%%%%%%%%%%%%%%%%%%%%%%%%%%%%%%%%%%%%%%%%%%%%%%%%%%%%%%%%%%%%%%%%%%%%%%%%%%%%%%%%%%%%%%%%%%%%%%%%%%%%%%%%
\section{Appell $F_1$ series and Gauss hypergeometric function} 
\label{sec:appendix1}

In this section, we provide an overview of the Appell $F_1$ series and the Gauss hypergeometric function~\cite{special_functions_book}. 

The Appell $F_1$ series is defined for $|x| < 1$ and $|y| < 1$ by
\begin{equation}
\label{eq:appell}
F_1(a, b_1, b_2, c;x,y) = \sum_{m=0}^\infty \sum_{n=0}^\infty \frac{(a)_{m+n}(b_1)_m(b_2)_n}{(c)_{m+n}m!n!}x^m y^n,
\end{equation}
where $(q)_n$ is the Pochhammer symbol. For values of $x$ and $y$ outside the range $|x| < 1$ and $|y| < 1$, the function $F_1$ can be extended through analytic continuation~\cite{bateman1953higher}. Such continuations can be achieved by manipulating integral representations, similar to the one in Eq.~(\ref{eq:picard}), where changing the integration variable can allow the expression of the original $F_1$ series through another $F_1$ series, e.g., see Eq.~(\ref{eq:rc_appel}). Such transformations enable the definition of the $F_1$ series for a broader range of $x$ and $y$.

The Gauss hypergeometric function is defined by the series
\begin{equation}
\label{eq:gauss}
{}_2 F_{1}[a, b, c; z]=\sum_{n=0}^\infty \frac{(a)_n (b)_n}{(c)_n} \frac{z^n}{n!},
\end{equation}
for $|z|<1$, and by analytic continuation elsewhere. 

The Appell $F_1$ series $F_1(a, b_1, b_2, c; x, y)$ degenerates to the Gauss hypergeometric function when $x=y$,
\begin{equation}
F_1(a, b_1, b_2, c; x, x)={}_2 F_{1}[a, b_1+b_2, c; x].
\end{equation}

%%%%%%%%%%%%%%%%%%%%%%%%%%%%%%%%%%%%%%%%%%%%%%%%%%%%%%%%%%%%%%%%%%%%%%%%%%%%%%%%%%%%%%%%%%%%%%%%%%%%%%%%%%%%%%%%%%%%%%%
\section{Tail of the contact distribution for any $\omega_1, \omega_2 \in [0,1)$} 
\label{sec:appendix2}

Here we establish that Eq.~(\ref{eq:h1approx}) in the main text holds true for any combination of $\omega_1, \omega_2 \in [0,1)$. To this end, we utilize the transformation given by Eq.~(1) in section~5.11 of Ref.~\cite{bateman1953higher}, which states that
\begin{align}
\label{eq:trans1}
\nonumber F_1 [a, b, b', c; x, y] & = (1-x)^{-b}(1-y)^{-b'}\\
& \times F_1 [c-a, b, b',c; \frac{x}{x-1}, \frac{y}{y-1}].
\end{align}
Applying this transformation to the $F_1$ function on the left-hand side of Eq.~(\ref{eq:h_1}), allows us to rewrite $h_1$ as
\begin{align}
\nonumber h_1 &= \omega_1^{t-1}\Big(\frac{1-\omega_2}{1-\omega_1}\Big)\\
& \times F_1 [1-T, 1-t, 1, 3; 1-\frac{1}{\omega_1}, \frac{\omega_2-\omega_1}{1-\omega_1}].
\label{eq:h_1app}
\end{align}
Now, using Eq.~(\ref{eq:asymptres}) with $a=1-T$, $b=1$, $\lambda=-t$, $b'=1$, $c=3$, $x=1-\frac{1}{\omega_1}$, and  $y=\frac{\omega_2-\omega_1}{1-\omega_1}$, we can write
\begin{widetext}
\begin{align}
\nonumber h_1 &= \omega_1^{t-1}\Big(\frac{1-\omega_2}{1-\omega_1}\Big) \sum_{n=0}^{m-1} \frac{(1-T)_n }{(3)_n}\Big(\frac{\omega_2-\omega_1}{1-\omega_1}\Big)^n {_2}F_1 [1-t, 1-T+n, 3+n; 1-\frac{1}{\omega_1}] + O\Big(\frac{1}{(-t)^{1-T+m}}\Big)\\
& = \Big(\frac{1-\omega_2}{1-\omega_1}\Big) \sum_{n=0}^{m-1} \frac{(1-T)_n }{(3)_n}\Big(\frac{\omega_2-\omega_1}{1-\omega_1}\Big)^n {_2}F_1 [2+T, 1-t, 3+n; 1-\omega_1] + O\Big(\frac{1}{(-t)^{1-T+m}}\Big).
\label{eq:h1expansionapp}
\end{align} 
\end{widetext}
The last equality follows from Pfaff's transformation (Eq.~(22) in section 2.1.4 of Ref.~\cite{bateman1953higher}), which states that
\begin{equation}
\label{eq:Pfaff2}
{}_2 F_{1}[a, b, c; z] = (1-z)^{-a} {}_2 F_{1}[a, c-b, c; \frac{z}{z-1}].
\end{equation}
We also utilized that ${}_2 F_{1}[a, b, c; z]={}_2 F_{1}[b, a, c; z]$, which follows from Eq.~(\ref{eq:gauss}).

\begin{figure}[!b]
\includegraphics[width=3in]{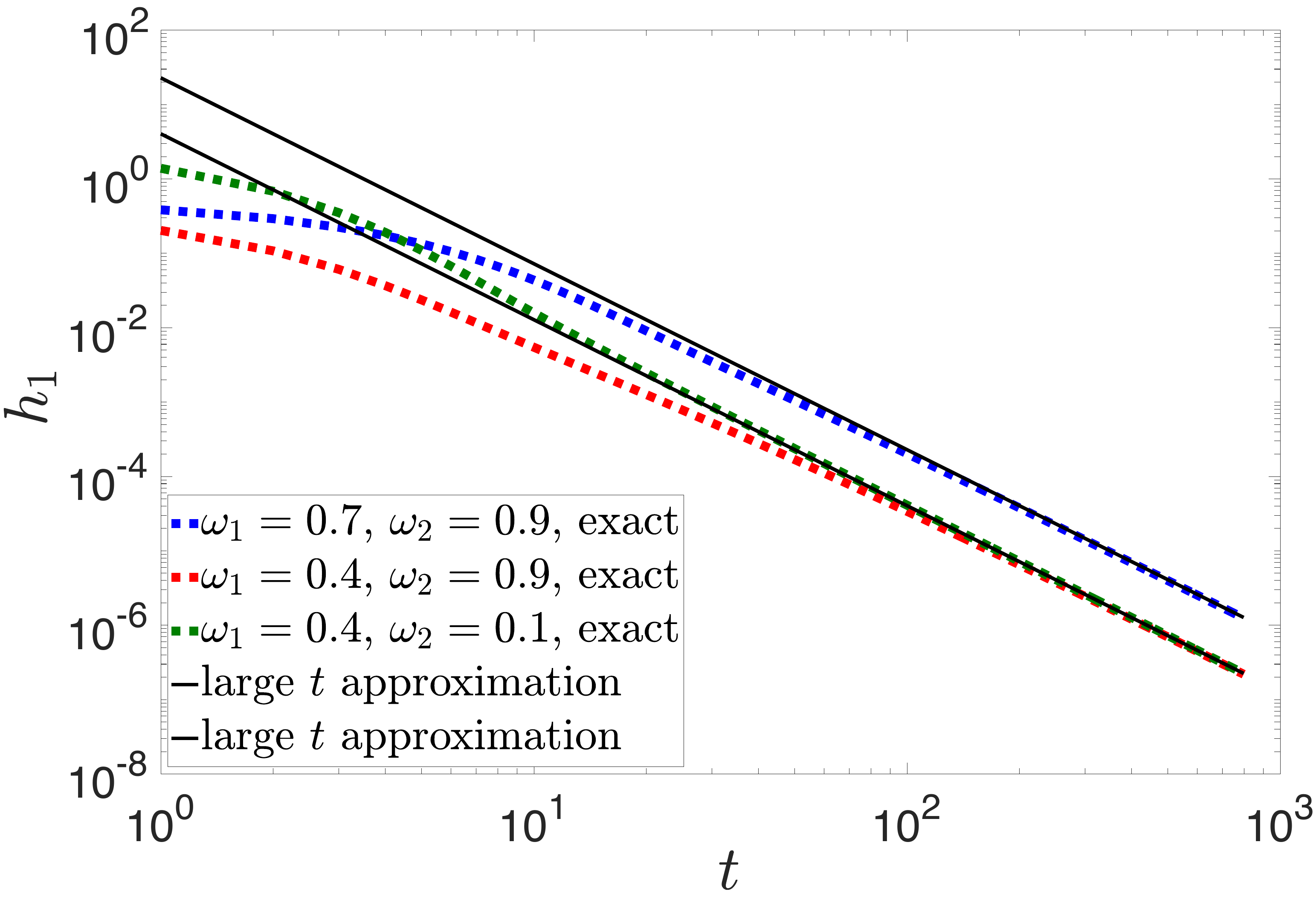}
\caption{
Function $h_1$ in Eq.~(\ref{eq:h_1app}) (dotted lines) vs. the approximation for large $t$ in Eq.~(\ref{eq:h1approx2_app}) (solid lines). Results are shown for different values of $\omega_1$ and $\omega_2$, while $T=0.5$. All axes use a logarithmic scale.
\label{fig:approx2_val}}
\end{figure}

Utilizing the asymptotic expansion for the hypergeometric function ${}_2 F_{1} [a, b, c; z]$ for $|b| \to \infty$, given by Eq.~(15) in section~2.3.2 of Ref.~\cite{bateman1953higher}, we can express the ${}_2 F_{1}$ function inside the sum in Eq.~(\ref{eq:h1expansionapp}), as
\begin{widetext}
\begin{align}
\nonumber {_2}F_1 [2+T, 1-t, 3+n; 1-\omega_1]&=\Bigg\{\frac{\Gamma{(3+n)}}{\Gamma{(1-T+n)}}\frac{(1-\omega_1)^{-(2+T)}}{(t-1)^{2+T}}+\frac{\Gamma{(3+n)}}{\Gamma{(2+T)}}\frac{e^{-(1-\omega_1)(t-1)}}{[(1-\omega_1)(1-t)]^{1-T+n}}\Bigg\}\\
&\times \Big[1+O\Big(\frac{1}{(1-\omega_1)(t-1)}\Big)\Big].
\label{eq:f1expansion}
\end{align}
\end{widetext}
At large $t$ the dominant term in Eq.~(\ref{eq:f1expansion}) is the first term inside the brackets, and we can write
\begin{widetext}
\begin{equation}
\label{eq:f1approxapp}
{_2}F_1 [2+T, 1-t, 3+n; 1-\omega_1] \approx \frac{\Gamma{(3+n)}}{\Gamma{(1-T+n)}}\frac{(1-\omega_1)^{-(2+T)}}{t^{2+T}}.
\end{equation}
\end{widetext}
Consequently, for large $t$ we can approximate Eq.~(\ref{eq:h1expansionapp}) as
\begin{widetext}
\begin{align}
\nonumber h_1 & \approx \Big(\frac{1-\omega_2}{1-\omega_1}\Big) \frac{(1-\omega_1)^{-(2+T)}}{t^{2+T}} \sum_{n=0}^{\infty} \frac{(1-T)_n }{(3)_n}\Big(\frac{\omega_2-\omega_1}{1-\omega_1}\Big)^n \frac{\Gamma{(3+n)}}{\Gamma{(1-T+n)}}\\
\nonumber &= \Big(\frac{1-\omega_2}{1-\omega_1}\Big) \frac{(1-\omega_1)^{-(2+T)}}{t^{2+T}} \frac{2}{\Gamma{(1-T)}} \sum_{n=0}^{\infty}\Big(\frac{\omega_2-\omega_1}{1-\omega_1}\Big)^n\\
& = \frac{2(1-\omega_1)^{-(2+T)}}{\Gamma{(1-T)}} \frac{1}{t^{2+T}}.
\label{eq:h1approx2_app}
\end{align} 
\end{widetext}
We see that the above analysis also leads to Eq.~(\ref{eq:h1approx}). We validate the analysis in Fig.~\ref{fig:approx2_val}.

\begin{figure}[!b]
\includegraphics[width=2.0in]{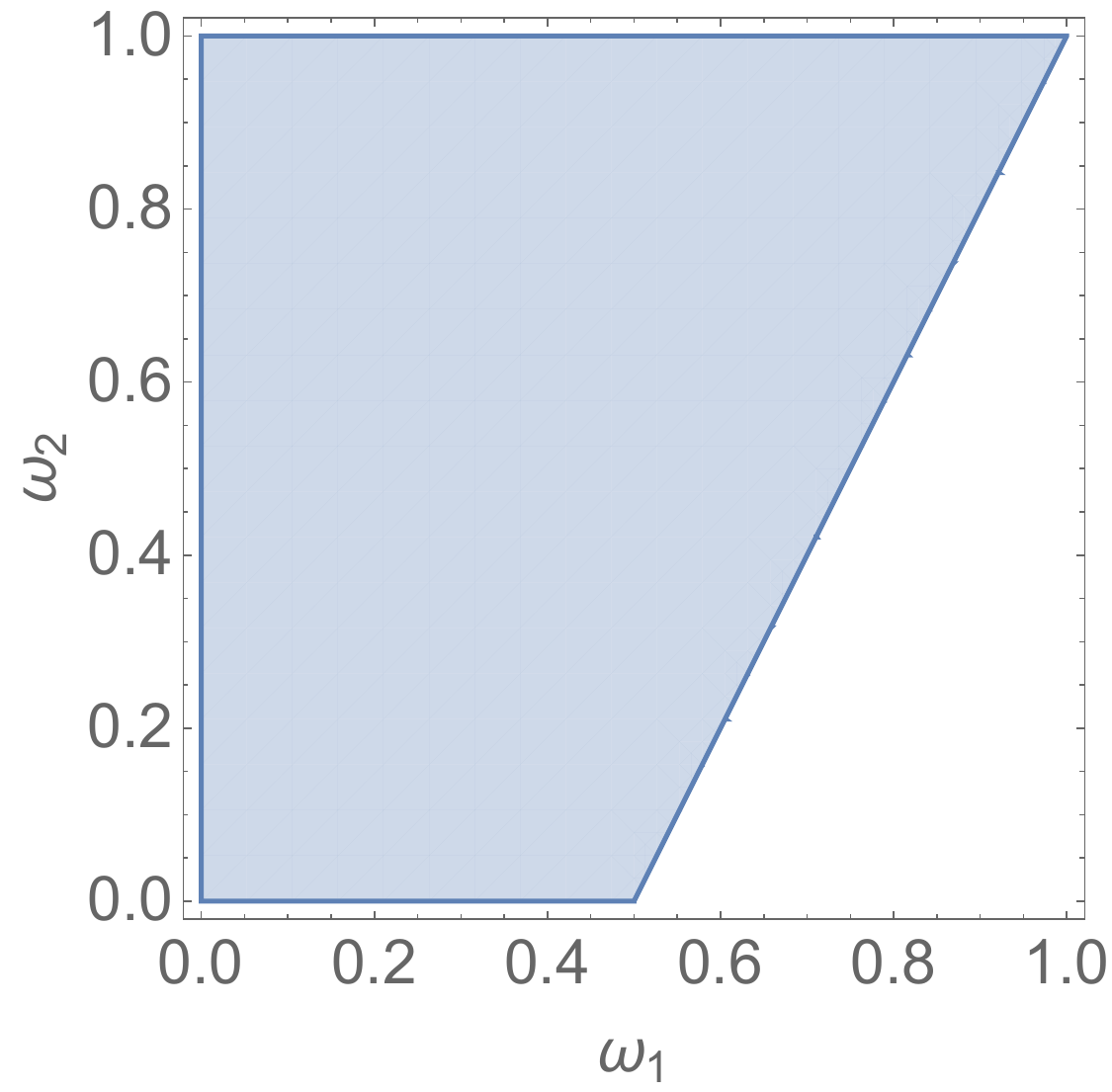}
\caption{Region $\mathcal{R}_2:=\{(\omega_1, \omega_2) \in \mathbb{R}^2 \ \mid \ |\frac{\omega_2-\omega_1}{1-\omega_1}| < 1\}$, shown as the blue-shaded area in the figure. In this region, Eq.~(\ref{eq:h1approx2_app}) holds.} 
\label{fig:region2}
\end{figure}

We note that in Eq.~(\ref{eq:h1approx2_app}), we let the summation run to infinity, since there is no single dominant term. The summation converges to $(1-\omega_1)/(1-\omega_2)$ when $|\frac{\omega_2-\omega_1}{1-\omega_1}| < 1$. This defines the region $\mathcal{R}_2$ of $\omega_1$ and $\omega_2$, depicted in Fig.~\ref{fig:region2}, for which the above analysis holds. The union of $\mathcal{R}_2$ with $\mathcal{R}_1$ in Fig.~\ref{fig:region1} covers the full range of $\omega_1, \omega_2 \in [0,1)$. Therefore, Eq.~(\ref{eq:h1approx}), and hence the scaling $P_{\textnormal{c}}(t) \propto 1/t^{2+T}$, hold for any  combination of $\omega_1$ and $\omega_2$.

\emph{Proving Eq.~(\ref{eq:f1approx}).} Equation~(\ref{eq:f1approx}) in the main text is obtained by using the same asymptotic expansion for the hypergeometric function ${}_2 F_{1} [a, b, c; z]$ for $|b| \to \infty$ as above (given by Eq.~(15) in section~2.3.2 of Ref.~\cite{bateman1953higher}). Specifically, utilizing this expansion, we can write
\begin{widetext}
\begin{align}
\label{eq:f1expansion2}
\nonumber {_2}F_1[2+T+n, 1+t, 3+n; 1-\frac{1}{\omega_1}] &= \Bigg\{\frac{\Gamma{(3+n)}}{\Gamma{(1-T)}}\frac{(1/\omega_1-1)^{-(2+T+n)}}{(t+1)^{2+T+n}}+\frac{\Gamma{(3+n)}}{\Gamma{(2+T+n)}}\frac{e^{-(1/\omega_1-1)(t+1)}}{[(1-1/\omega_1)(t+1)]^{1-T}}\Bigg\}\\
&\times \Big[1+O\Big(\frac{1}{(1/\omega_1-1)(t+1)}\Big)\Big].
\end{align}
\end{widetext} 
At large $t$, the dominant term in the above relation is the first term inside the brackets. Utilizing also that ${}_2 F_{1}[a, b, c; z]={}_2 F_{1}[b, a, c; z]$, we can write
\begin{widetext}
\begin{equation}
{_2}F_1[1+t, 2+T+n, 3+n; 1-\frac{1}{\omega_1}] \approx \frac{\Gamma{(3+n)(1/\omega_1-1)^{-(2+T+n)}}}{\Gamma{(1-T)}}\frac{1}{t^{2+T+n}}.
\end{equation}
\end{widetext}

%%%%%%%%%%%%%%%%%%%%%%%%%%%%%%%%%%%%%%%%%%%%%%%%%%%%%%%%%%%%%%%%%%%%%%%%%%%%%%%%%%%%%%%%%%%%%%%%%%%%%%%%%%%%%%%%%%%%%%%
%

\end{document}